\documentclass[english,aps,prl,twocolumn]{revtex4-2}
\usepackage[T1]{fontenc}
\usepackage[latin9]{inputenc}
\usepackage{babel}
\usepackage{textcomp}
\usepackage{mathrsfs}
\usepackage{mathtools}
\usepackage{amsmath}
\usepackage{amssymb}
\usepackage{nicefrac}
\usepackage{graphicx}
\usepackage{natbib}

\makeatletter
\usepackage{physics}
\usepackage{multirow}

\usepackage{hyperref}
\makeatother

\begin{document}
	
\title{A Hamiltonian Mechanics Framework for Charge Particle Optics in Straight
and Curved Systems}
\author{\noindent F. Kern}
\affiliation{Institute of Solid State Research, IFW Dresden, 01069 Dresden, Germany}

\author{\noindent J. Krehl}
\affiliation{Institute of Solid State Research, IFW Dresden, 01069 Dresden, Germany}

\author{\noindent A. Thampi}
\affiliation{Institute of Solid State Research, IFW Dresden, 01069 Dresden, Germany}

\author{\noindent A. Lubk}
\affiliation{Institute of Solid State Research, IFW Dresden, 01069 Dresden, Germany}

\begin{abstract}
Charged particle optics, the description of particle trajectories
in the vicinity of some optical axis, describe the imaging properties
of particle optics devices. Here, we present a complete and compact
description of charged particle optics employing perturbative expansion
of Hamiltonian mechanics. The derived framework allows the straightforward
computation of transversal and longitudinal (chromatic) properties
of static and dynamic optical devices with straight and curved optical
axes. It furthermore gives rise to geometric integration schemes preserving
the symplectic phase space structure and pertaining Lagrange invariants,
which may be employed to derive analytic approximations of aberration
coefficients and efficient numerical trajectory solvers.
\end{abstract}

\maketitle

\section{Introduction}

Charged particle optical (CPO) systems such as particle accelerators,
electron or ion microscopes, focussed ion beam instruments, or lithography
machines rely on tailored electromagnetic field configurations to
realize optical imaging with charged particles. The goal is to create
a controllable and well-defined mapping between entrance and exit
plane of the device. This can be typically described well within the
framework of (relativistic) classical mechanics as quantum effects
(e.g., spin-orbit coupling) are generally small. Moreover, one does
not need the general description of all possible particle trajectories
in a given field configuration; those close to the optical axis are
generally sufficient. For instance, the paraxial approximation, which
describes first order deviations of trajectories with respect to the
optical axis already entails the principal properties of the optical
system. Unfortunately, the large numerical apertures and beam diameters
used in modern CPO devices typically requires consideration of trajectories
beyond the paraxial, i.e., linear, limit. These trajectories determine
the aberrations limiting the optical performance of the device (e.g.,
spatial resolution, beam collimation or energy filter transparency).
Although modern computational methods allow numerical evaluation of
arbitrary trajectories at high speed and accuracy, perturbation methods
computing path deviations in a semianalytical way are indispensible
tools as they provide categorical properties, e.g., symmetries canceling
out certain aberrations, and parametric models, e.g., for refining
field configurations. The most common analytical perturbation methods
are the trajectory method, the eikonal method and the Lie algebra
method. The trajectory method iteratively solves Newton's equations
of motion \citep{Scherzer1933}, the eikonal method \citep{Glaser1933,Sturrock1951,Hawkes(1996),Rose2004}
iterates Hamiltonian characteristic functions (here called the eikonal)
with respect to trajectory deviations, and the Lie algebra method
\citep{Dragt1982,Dragt1987,Dragt1988} is based on the canonical
perturbation method in phase space \citep{Cary1977,Cary1981}. A comprehensive
comparison of these methods in electron optics has been carried out
by T. Radlicka \citep{Radlicka2009}.

In spite of the longstanding and successful use of the above approaches,
they also have some shortcomings, which, e.g., prevent a compact description
and complicate the treatment of curvilinear systems. Indeed, chromatic
aberrations are frequently derived by a separate variation of the
trajectories with respect to the particles energy and in devices with
with a curved optical axis curved coordinate systems (e.g., defined
by moving Frenet trihedrals) are employed (e.g., \citep{Hawkes(1988)}).
Both approaches (although perfectly viable from a theoretical point
of view) come as auxiliary modifications, frequently leading to significant
mathematical and computational complexity. We therefore provide for
yet another perturbation approach in the following, which leads to
very compact and straightforward expressions irrespective of the particular
shape of the CPO system, the fields, considered deviation parameters
or the chosen coordinate system.

Our approach is rooted in Hamiltonian mechanics exploiting the perturbation
framework leading to the so-called Jacobi variational equation (JVE)
in the first order \citep{Frankel2004}. The approach naturally makes
use of the symplectic structure of classical mechanics (i.e., conservation
of phase space), allowing, for example, the use of efficient geometrical
integration schemes in the calculation of optical properties from
given field configurations. In that, the approach shares a number
of similarities with the Lie algebraic methods \citep{Dragt1976,Dragt1982},
indeed it may be regarded as a compactified version thereof.

In the following, we will shortly recapitulate the basics of Hamiltonian
mechanics for charged particles including perturbation series expansion
of the Hamiltonian equations of motion. We then explicitly carry out
the linear perturbation for electric and magnetic fields in both the
non-relativistic and relativistic regime and discuss the structure
and general properties of the ensuing paraxial equations. In the subsequent
section higher-order expansions are demonstrated. In the Appendix
we discuss some of the most important CPO elements, namely magnetic
quadrupole, round magnetic lens, Wien filter and sector magnet, to
illustrate the developed formalism.

\section{Fundamentals of Hamiltonian Mechanics}

To describe the motion of a charged particle in an electromagnetic
field we employ Hamiltonian mechanics, which operates in phase space
endowed with phase space coordinates for position $\boldsymbol{r}$
and canonical momentum $\boldsymbol{p}$. The non-relativistic Hamiltonian,
corresponding to the total energy of the system, reads
\begin{equation}
\mathcal{H}\left(\boldsymbol{r},\boldsymbol{p},t\right)=\frac{\left(\boldsymbol{p}-q\boldsymbol{A}\left(\boldsymbol{r},t\right)\right)^{2}}{2m}+q\Phi\left(\boldsymbol{r},t\right)
\end{equation}
Here, $m$, $q$, $\boldsymbol{A}$, and $\Phi$ denote the particle's
mass, charge, as well as the magnetic vector potential and electric
potential, respectively. As charged particle optics frequently deals
with relativistic particles we also note the relativistic Hamiltonian
\begin{align}
{\displaystyle \mathcal{H}} & =\sqrt{m^{2}c^{4}+\left(\mathbf{p}-q\mathbf{A}\right)^{2}c^{2}}+q\Phi\\
 & =\gamma mc^{2}+q\Phi\,,\nonumber 
\end{align}
where we omitted the coordinate arguments to shorten notation. In
the following derivation, however, we resort to the non-relativistic
expressions for simplicity. Relevant relativistic expressions will
be given at the end of the next section. Hamilton's equations of motion
read
\begin{equation}
\frac{d\boldsymbol{x}}{dt}=\begin{pmatrix}\frac{d\boldsymbol{r}}{dt}\\
\frac{d\boldsymbol{p}}{dt}
\end{pmatrix}=\left\{ \boldsymbol{x},\mathcal{H}\right\} =\begin{pmatrix}\frac{\partial\mathcal{H}}{\partial\boldsymbol{p}}\\
-\frac{\partial\mathcal{H}}{\partial\boldsymbol{r}}
\end{pmatrix}=\boldsymbol{X}_{\boldsymbol{x}\left(t\right)}\,,\label{eq:HEOM}
\end{equation}
where we introduced typically 6-dimensional phase space coordinates
$\boldsymbol{x}\coloneqq\left(\boldsymbol{r},\boldsymbol{p}\right)^{T}$,
the Poisson bracket $\left\{ \cdot,\cdot\right\} $, and the Hamiltonian
vector field
\begin{equation}
\boldsymbol{X}_{\boldsymbol{x}\left(t\right)}\coloneqq\begin{pmatrix}\frac{\boldsymbol{p}-q\boldsymbol{A}}{m}\\
\frac{q}{m}\frac{\partial A_{i}}{\partial\boldsymbol{r}}\left(p_{i}-qA_{i}\right)-q\frac{\partial\Phi}{\partial\boldsymbol{r}}
\end{pmatrix}\,.\label{eq:Ham_vec_field}
\end{equation}
Here and henceforth, Einstein summation convention is employed.

In CPO we describe the motion of a charged particle in the vicinity
of some given optical axis (also referred to as the design trajectory),
which is an exact solution to the full equations of motion. That can
be a straight optical axis as, e.g., the symmetry axis of a round
magnetic lens, or a curved axis as, e.g., a circular reference trajectory
in a sector magnet. The Lie algebraic methods use the second equality
in (\ref{eq:HEOM}) as starting point for an iterative solution facilitated
by a Taylor expansion of the Hamiltonian around the the optical axis.
This formalism involves the use of some operator algebra and canonical
transformations to the design trajectory. In the following we essentially
present a shortcut yielding the same results based on directly considering
the variation of particle trajectory $\delta\boldsymbol{x}$ around
the optical axis trajectory $\boldsymbol{x}\left(t\right)$ in phase
space
\begin{equation}
\boldsymbol{y}\left(t\right)=\boldsymbol{x}\left(t\right)+\delta\boldsymbol{x}\left(t\right)\,.
\end{equation}
Hamilton's equations of motion for such trajectories read
\begin{equation}
\frac{d\left(\boldsymbol{x}+\delta\boldsymbol{x}\right)}{dt}=\boldsymbol{X}_{\boldsymbol{x}+\delta\boldsymbol{x}}
\end{equation}
yielding the following difference to the optical axis trajectory
\begin{align}
\frac{d\left(\boldsymbol{x}+\delta\boldsymbol{x}\right)}{dt}-\frac{d\boldsymbol{x}}{dt} & =\frac{d\left(\delta\boldsymbol{x}\right)}{dt}\label{eq:traj_diff}\\
 & =\boldsymbol{X}_{\boldsymbol{x}+\delta\boldsymbol{x}}-\boldsymbol{X}_{\boldsymbol{x}}\nonumber 
\end{align}
We now expand the deviation into a perturbation series, $\delta\boldsymbol{x}=\sum_{n}\varepsilon^{n}\delta\boldsymbol{x}^{\left(n\right)}$,
with an artificial smallness parameter $\varepsilon=1$, indicating
the smallness of the deviation vector $\delta\boldsymbol{x}^{\left(n\right)}$
and develop the right hand side of (\ref{eq:traj_diff}) into a multidimensional
Taylor series
\begin{align}
\frac{d\left(\sum_{n}\varepsilon^{n}\delta x_{i}^{(n)}\right)}{dt} & =\left(\frac{\partial X_{i}}{\partial x_{j}}\right)_{\boldsymbol{x}\left(t\right)}\sum_{n}\varepsilon^{n}\delta x_{j}^{\left(n\right)}\label{eq:perturb}\\
 & +\frac{1}{2}\left(\frac{\partial^{2}X_{i}}{\partial x_{j_{1}}\partial x_{j_{2}}}\right)_{\boldsymbol{x}\left(t\right)}\sum_{n_{1}}\varepsilon^{n_{1}}\delta x_{j_{1}}^{\left(n_{1}\right)}\sum_{n_{2}}\varepsilon^{n_{2}}\delta x_{j_{2}}^{\left(n_{2}\right)}\nonumber \\
 & +\ldots\nonumber \\
 & =\sum_{n=1}^{\infty}\frac{1}{n!}\sum_{\left|j\right|=n}\begin{pmatrix}n\\
j
\end{pmatrix}\left(D^{j}X_{i}\right)_{\boldsymbol{x}\left(t\right)}\left(\sum_{m}\varepsilon^{m}\delta x^{\left(m\right)}\right)^{j}\,.\nonumber 
\end{align}
In the last line we employed multiindex notation $j=\left(j_{1},...,j_{6}\right)$
with $D^{j}=\frac{\partial^{\left|j\right|}}{\partial x_{1}^{j_{1}}...\partial x_{6}^{j_{6}}}$
and $\begin{pmatrix}n\\
j
\end{pmatrix}=\frac{n!}{\prod_{i=1}^{6}j_{i}!}$. Equating terms of the same power of $\varepsilon$ gives rise to
a hierarchical system of differential equations defining deviations
in the vicinity of some design trajectory (optical axis). Here, the
initial values for the $\delta\boldsymbol{x}^{\left(n>1\right)}$
deviations (i.e., aberrations) will be deliberately set to zero (i.e.,
starting conditions of trajectories are completely absorbed by Gaussian
trajectory), which greatly facilitates a solution. Subsequently, we
discuss solutions to this hierarchical system starting with the first
order.

\section{Paraxial Optics}

Paraxial optics, which is also referred to as Gaussian or first order
optics, is an approximation to general charged particle optics, in
which one considers only those trajectories, which are very close
to the optical axis (which, by definition, is itself a valid trajectory).
In this limit the general equations of motions simplify considerably,
which allows, e.g., to set up linear relationships (so-called transfer
matrices) between the deviations in position and angle with respect
to the optical axis at different points along the optical axis. This
implies that trajectories emanating from the same object point all
intersect in an image point (which can be real, virtual, or at infinity,
however). Deviations from these Gaussian trajectories are referred
to as aberrations. Consequently, paraxial optics describes a CPO system
completely if no aberrations are present.

Gaussian optics serves as the starting and reference point for the
description of any particle optics system, from which design principles
and further considerations are derived. In that, one typically identifies
an optical axis in the first step (e.g., analytically using symmetry
arguments or numerically by ray tracing) and computes a number of
important optical characteristics within the paraxial approximation
in a second step. The latter includes, e.g., the effective focal length,
location of the principal planes and magnification. The final step
consist of an analysis of the deviations from the Gaussian behaviour,
i.e., aberrations, where the Gaussian trajectories are again very
important for the evaluation of the perturbation series (see next
section). Because of the distinguished significance of paraxial optics,
we give a more detailed account of it in the following, which also
serves us to introduce relevant notation.

\begin{figure}
\includegraphics[width=1\columnwidth]{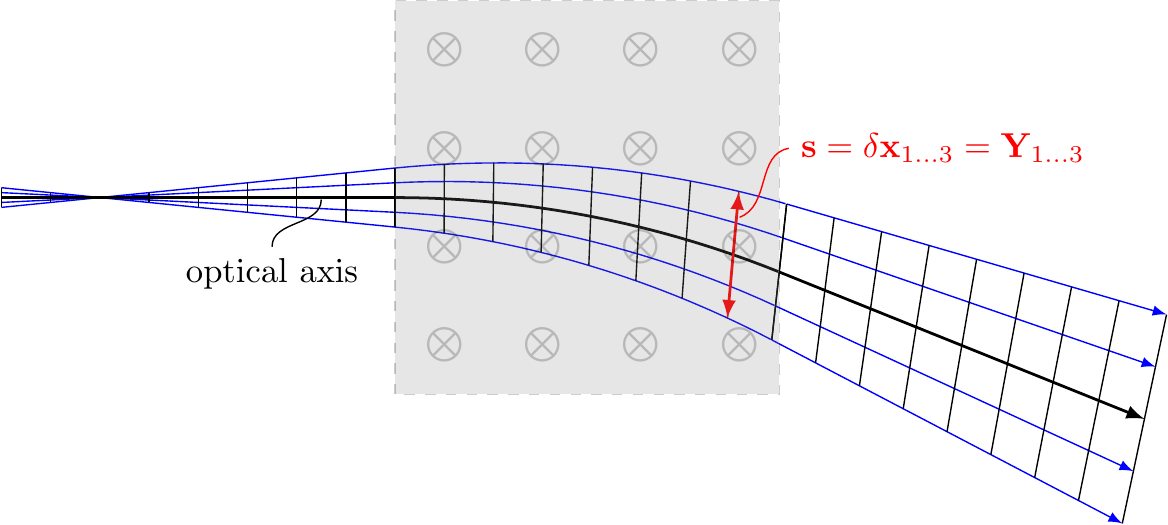}\caption{\label{fig:Paraxial-trajectories}Paraxial trajectories in a constant
transverse magnetic field of a magnetic deflector with sharp cut-offs.
The thick red arrows denote the spatial deviation $\boldsymbol{s}\protect\coloneqq\delta\boldsymbol{x}_{1\ldots3}=\boldsymbol{Y}_{1\ldots3}$
from the black reference trajectory (optical axis). Note that both
transversal and longitudinal deviations to the reference trajectory
are present in this example.}
\end{figure}

Keeping only lowest order in the Taylor expansion (\ref{eq:perturb})
pertains to the paraxial case and yields the so-called Jacobi variation
equation (JVE) in coordinate expression
\begin{align}
\frac{d\left(\delta x_{i}\left(t\right)\right)}{dt} & \approx\left(\frac{\partial X_{i}}{\partial x_{j}}\right)_{\boldsymbol{x}\left(t\right)}\delta x_{j}\left(t\right)\label{eq:Jac_var_eq}\\
 & =\chi_{ij}\left(t\right)\delta x_{j}\left(t\right)\,,\nonumber 
\end{align}
where we introduced the short-hand notation $\chi$$\left(t\right)$
for the Jacobian of the Hamiltonian flow. A solution of these equations
in a homogeneous magnetic field with sharp cut-offs is displayed in
Fig. \ref{fig:Paraxial-trajectories}. Note that the last equation
is exact if second and higher order derivatives of the Hamiltonian
vector field are zero. 

We now introduce the notion of the Lie derivative (see Appendix \ref{sec:Lie-Derivative})
by defining the path deviation vector $\boldsymbol{Y}\coloneqq\delta\boldsymbol{x}$,
which allows to reexpress the JVE (\ref{eq:Jac_var_eq}) as a disappearance
of the Lie derivative of the deviation vector $\boldsymbol{Y}$ along
the optical axis (defined by vector field $\boldsymbol{X}$) $\mathcal{L}_{\boldsymbol{X}}\boldsymbol{Y}=0$
(see Appendix \ref{sec:Lie-Derivative}). In other words, the deviation
vectors $\boldsymbol{Y}$ form an invariant vector field along the
optical axis, which has the following consequence. Inserting any two
invariant vector fields $\boldsymbol{Y}$, $\boldsymbol{Z}$ (trajectories)
as arguments into Liouville's theorem yields 
\begin{equation}
0=\mathcal{L}_{\boldsymbol{X}}\omega\left(\boldsymbol{Y},\boldsymbol{Z}\right)=\frac{d}{dt}\left[\omega\left(\boldsymbol{Y},\boldsymbol{Z}\right)\right]\,,
\end{equation}
meaning that the phase space area spanned by $\boldsymbol{Y}$ and
$\boldsymbol{Z}$ (measured by the Poincare 2-form $\omega$) along
the optical axis is preserved. The conservation of transverse phase
space areas (i.e., $\boldsymbol{Y}_{\bot}$and $\boldsymbol{Z}_{\bot}$
perpendicular to optical axis) in the object and image plane of anastigmatic
optical elements (e.g., round lens) is referred to as the Lagrange-Helmholtz
invariant. Similar Lagrange invariants follow for other elements such
as energy filters (see Fig. \ref{fig:Paraxial-trajectories} for an
energy-dispersive deflector). From now on we shall always consider
CPO as the transport of some (deviation) vector along the optical
axis.

Inserting the Hamiltonian vector field (\ref{eq:Ham_vec_field}) into
(\ref{eq:Jac_var_eq}), we obtain the coordinate expression for deviation
vector motion in phase space\begin{widetext}
\begin{align}
\frac{dY_{i}}{dt} & =\chi_{ij}Y_{j}\label{eq:Lie_coord}\\
 & =\frac{q}{m}\left(\begin{array}{c|c}
-\frac{\partial A_{i}}{\partial r_{j}} & \frac{1}{q}\delta_{ij-3}\\
\hline \frac{\partial^{2}A_{k}}{\partial r_{i-3}\partial r_{j}}\left(p_{k}-qA_{k}\right)-q\frac{\partial A_{k}}{\partial r_{i-3}}\frac{\partial A_{k}}{\partial r_{j}}-m\frac{\partial^{2}\Phi}{\partial r_{i-3}\partial r_{j}} & \frac{\partial A_{j-3}}{\partial r_{i-3}}
\end{array}\right)_{\boldsymbol{x}\left(t\right)}Y_{j}\nonumber 
\end{align}
\end{widetext}We can readily verify that $J\chi$ with the skew-symmetric
$6\times6$ matrix ($I_{3}$ denotes the $3\times3$ identity matrix)
\begin{equation}
J=\begin{bmatrix}0 & I_{3}\\
-I_{3} & 0
\end{bmatrix}
\end{equation}
is symmetric, which is the defining property for a Hamiltonian matrix
that form the symplectic Lie algebra $\mathrm{\boldsymbol{sp}}\left(6,\mathbb{R}\right)$.
The corresponding exponential map including multiplications generate
the symplectic group $\mathrm{Sp}\left(6,\mathbb{R}\right)$, which
is another way of saying that deviation vectors governed by the JVE
preserve phase space areas (see further below).

The above Lie derivative of a deviation vector in phase space is not
gauge invariant. The equations of motion of the position vector ($\boldsymbol{s}=\boldsymbol{Y}_{1\ldots3}$)
derived by a second derivation of the first three entries of (\ref{eq:Lie_coord})
\begin{align}
m\frac{d^{2}s_{i}}{dt^{2}} & =-q\frac{\partial A_{i}}{\partial r_{j}}\frac{ds_{j}}{dt}-q\frac{\partial^{2}A_{i}}{\partial r_{j}\partial r_{k}}\frac{dr_{k}}{dt}s_{j}+\frac{dY_{i+3}}{dt}\label{eq:EoM}\\
 & =-qF_{ij}\frac{ds_{j}}{dt}-q\frac{\partial F_{ij}}{\partial r_{j}}\frac{dr_{k}}{dt}s_{j}+q\frac{\partial E_{i}}{\partial r_{j}}s_{j}\,,\nonumber 
\end{align}
however, contain only physical fields, i.e., the electric field $\boldsymbol{E}$
and the magnetic fields $\boldsymbol{B}$ (the latter as spatial components
of the electromagnetic tensor $\boldsymbol{F}$) which are gauge invariant.
In systems with straight optical axis, the transverse part (perpendicular
to the optical axis) of these equations of motion are consistent with
the usual paraxial approximation of the trajectory method derived
from truncating the mulitpolar expansion of the electromagnetic fields
around the optical axis to linear terms (e.g., \citep{Hawkes(1996)})
with one notable exception: Being explicitly constructed as differences
to a design trajectory the above paraxial equations do not contain
any axial acceleration potential as this is already absorbed into
the reference trajectory (i.e., optical axis).

The above JVE (\ref{eq:Lie_coord}) and the ensuing paraxial Newtonian
equations of motion (\ref{eq:EoM}) treat both transversal and longitudinal
deviations as well as straight and curvilinear systems on the same
footing. Both describe paraxial optics completely and can be efficiently
solved by numerical (step) solvers, which is the standard procedure,
if the electric and magnetic fields vary along the optical axis. Solving
the second order differential equation (\ref{eq:EoM}) does not warrant
gauge considerations and requires several integrations to obtain all
fundamental solutions completely describing a given system. Integrating
the JVE, on the other hand, directly yields the transfer matrix from
one plane to another, however, gauge needs to be considered generally
(see below). The latter argument may be turned around, however, as
gauge freedom can be sometimes exploited (A) to simplify the solution
of the coupled systems of first-order differential equations (\ref{eq:Lie_coord})
and (B) to simplify the relation between kinetic and canonical momentum
in particular planes of interest (e.g., object and image plane). One
particular useful trick is to fix the gauge such to render $\chi\left(t\right)$
time-independent (i.e., constant along the optical axis), which leads
to a particularly simple integration of (\ref{eq:Lie_coord}). Moreover,
conservation of phase space is an additional structure available in
the JVE, which can be incorporated into efficient approximation schemes
(i.e., geometrical integration, see below). These arguments carry
over to the perturbation schemes rooted either in Newtonian equations
of motion (i.e.the trajectory method) or Hamiltonian (next section)
and Hamilton-Jacobi mechanics (i.e. the Lie algebra method).

To illustrate these aspects and to further the analytical treatment
of the paraxial and higher order trajectory equations we elaborate
on the solution of the JVE in the following. The formal solution to
Eq. (\ref{eq:Lie_coord}) is obtained by the time-ordered (symplectic)
exponential map

\begin{equation}
Y_{i}\left(t\right)=\underset{\mathcal{M}_{ij}}{\underbrace{\mathcal{T}\exp\left\{ \int_{0}^{t}\chi_{ij}\left(t_{1}\right)\dd t_{1}\right\} }}Y_{j}\left(0\right)\,,\label{eq:Transfer_Matrix}
\end{equation}
where $\mathcal{T}$ is the time ordering operator. Thereby, the defined
matrix $\mathcal{M}$ allows to compute the deviation vector $\boldsymbol{Y}\left(t\left(l\right)\right)$
of some trajectory at some time, defined by the arc length $l=\int_{0}^{t}v_{0}\left(t_{1}\right)\dd t_{1}$
of the optical axis given its initial phase space coordinates $\boldsymbol{Y}\left(0\right)$.
As $\mathcal{M}$ is symplectic its inverse can be readily obtained
from $\mathcal{M}^{-1}=-J\mathcal{M}^{T}J$, which is again in the
symplectic group $\mathrm{Sp}\left(6,\mathbb{R}\right)$. We will
see further below that the last property is a key to a straightforward
evaluation of aberrations.

Here it is important to note that the locus of the spatial part of
$\boldsymbol{Y}\left(t\right)$ at some fixed time $t$ is typically
not confined to a single (image) plane even if we restrict the spatial
part of $\boldsymbol{Y}\left(0\right)$ to a plane (i.e., all trajectories
start in some (object) plane, see Fig. \ref{fig:Paraxial-trajectories}
for an example). In other words, there generally is a spatial longitudinal
deviation between the spatial end point of the deviation vector and
some predefined image plane intersected by the optical axis at $l$
(which is often but not necessarily perpendicular to the optical axis).
This in turn leads to a positive or negative time lag $\delta t$
for the corresponding trajectory to reach the image plane, which is
of first order in (i.e., depends linearly on) the initial deviation
vector $\boldsymbol{Y}\left(0\right)$ of the trajectory. This leads
to additional first order transversal deviations $\boldsymbol{Y}^{\left(t\right)}$
in the image plane that must be considered in the paraxial limit,
if a non-zero deviation vector is produced by the additional time
required to propagate the particle along the optical axis to the image
plane, i.e., 
\begin{equation}
\boldsymbol{Y}^{\left(t\right)}\left(t\right)\approx\boldsymbol{X}_{\left(t\right)}\delta t\,.\label{eq:par_timelag}
\end{equation}
Higher order propagation effects (i.e., stemming from the additional
propagation of the deviation vector via $\mathcal{M}\left(\delta t\right)$
in (\ref{eq:Transfer_Matrix})), on the other hand, may be neglected
in the paraxial regime because these would be totally of second order
in the initial beam coordinates as the time lag itself is already
of first order. Consequently, the additional correction due to longitudinal
deviations is important only when considering final planes located
at a curved optical axis and we discuss one particular example, the
sector magnet, in Appendix \ref{subsec:Sector-Magnet}. Note, however,
that this correction vanishes, if image planes outside of the fields
are considered. Here, the longitudinal transport along the optical
axis does not introduce any additional first order effects as the
zeroth order trajectory (optical axis) is straight and hence $\boldsymbol{Y}{}_{xy}^{\left(t\right)}\left(t\right)=0$.
This condition can be also achieved ``artificially'' by rectifying
a curved optical axis in curvilinear coordinates, which is the key
argument for their use in literature. Here, we argue, however, that
the paraxial time lag correction (\ref{eq:par_timelag}), if necessary
at all (i.e., if considering planes at bend sections of the optical
axis), is easy to compute, hence not increasing complexity in comparison
to those coming with curvilinear coordinates. 

In case of elements with straight optical axis, the transverse subspace
of $\mathcal{M}$ corresponds to the transfer matrices of paraxial
optics if $\boldsymbol{A}_{\perp}=0$ in the initial and final plane.
Occasionally, gauge freedom can be exploited to achieve that condition.
In the general case, however, a transformation $\mathcal{G}$ sending
kinetic to canonical momentum 
\begin{align}
\mathcal{G} & :\delta\boldsymbol{p}\left(t\right)\rightarrow m\delta\boldsymbol{v}\left(t\right)=\mathcal{G}\left(t\right)\circ\delta\boldsymbol{p}\left(t\right),\\
 & M\rightarrow M=\mathcal{G}\left(t\right)\circ\mathcal{M}\circ\mathcal{G}^{-1}\left(0\right)\nonumber 
\end{align}
is required ($\circ$ denotes the application of that transformation)
to obtain the gauge invariant transfer matrix for configurations space
(i.e., consisting of $\boldsymbol{r}$ and kinetic momentum)
\begin{equation}
\boldsymbol{Y}^{\mathrm{kin}}\left(t\right)=M\boldsymbol{Y}^{\mathrm{kin}}\left(0\right)\,.
\end{equation}
A particular simple situation is encountered if $\mathcal{G}$ is
linear in the position vector, i.e,. may be represented by a matrix
$G$ (see discussion of the Wien filter in Appendix \ref{sec:Examples}
for an example)
\begin{equation}
M=G\left(t\right)\mathcal{M}G^{-1}\left(0\right)\,.\label{eq:sim_trafo}
\end{equation}
Eq. (\ref{eq:Transfer_Matrix}) generally does not admit closed expressions
for the transfer matrix and requires using numerical integrators (e.g.,
directly applied to (\ref{eq:Lie_coord})). Note, however, that closed
solutions exist for a number of special cases; moreover, series expansions
may be used to obtain analytical approximations of increasing accuracy.
Solving (\ref{eq:Transfer_Matrix}) is straightforward if $\chi\left(t\right)$
commutate for all times $t$ (which includes time-independent $\chi$),
yielding
\begin{align}
\mathcal{M} & =\exp\left(\int_{0}^{t}\chi\left(t_{1}\right)\dd t_{1}\right)\,.\label{eq:weak_M}
\end{align}
Indeed, increasingly better approximation to general transfer matrices
may be obtained from the Magnus expansion\citep{Magnus1954}
\begin{equation}
{\displaystyle \mathcal{M}=\exp\left(\sum_{k=1}^{\infty}\Omega_{k}\right)\,,}\label{eq:Magnus}
\end{equation}
with the first two terms ($k=1,2$) explicitly reading
\begin{equation}
{\displaystyle \begin{aligned}\Omega_{1}(t) & =\int_{0}^{t}\chi(t_{1})\dd t_{1}\\
\Omega_{2}(t) & =\frac{1}{2}\int_{0}^{t}\dd t_{1}\int_{0}^{t_{1}}\dd t_{2}[\chi(t_{1}),\chi(t_{2})]\,.
\end{aligned}
}
\end{equation}
Here, $\left[\cdot,\cdot\right]$ denotes the commutator. Notably,
the Magnus expansion conserves phase space and hence Lagrange invariants
at every order of the approximation (i.e., truncation of (\ref{eq:Magnus})),
a property that is not shared by the class of multistep integrators
or Runge-Kutta methods. In other words, it falls in the class of geometric
integrators\citep{Leimkuhler2005}, rendering it a useful tool to
describe the paraxial optics of general systems with fields varying
along the optical axis (see discussion of the round lens in Appendix
\ref{sec:Examples} for an exemplary application of Magnus expansion).

We finally also note the fully relativistic expressions for the Hamiltonian
vector field
\begin{align}
\boldsymbol{X} & =\begin{pmatrix}{\displaystyle \frac{\mathbf{p}-q\mathbf{A}}{\sqrt{m^{2}+\frac{1}{c^{2}}\left(\mathbf{p}-q\mathbf{A}\right)^{2}}}}\\
{\displaystyle q\frac{\partial A^{i}}{\partial\boldsymbol{r}}{\displaystyle \frac{p_{i}-qA_{i}}{\sqrt{m^{2}+\frac{1}{c^{2}}\left(\mathbf{p}-q\mathbf{A}\right)^{2}}}}-q\frac{\partial\Phi}{\partial\boldsymbol{r}}}
\end{pmatrix}\\
 & =\begin{pmatrix}{\displaystyle \frac{\mathbf{p}-q\mathbf{A}}{\gamma m}}\\
{\displaystyle q\frac{\partial A_{i}}{\partial\boldsymbol{r}}{\displaystyle \frac{p_{i}-qA_{i}}{\gamma m}}-q\frac{\partial\Phi}{\partial\boldsymbol{r}}}
\end{pmatrix}\,,\nonumber 
\end{align}
and the ensuing JVE\begin{widetext}
\begin{equation}
\frac{dY_{i}}{dt}=\frac{q}{\gamma m}\left(\begin{array}{c|c}
-\frac{\partial A_{i}}{\partial r_{j}}+\frac{\partial A_{k}}{\partial r_{j}}\frac{p_{k}-qA_{k}}{c^{2}\left(\gamma m\right)^{2}}\left(p_{i}-qA_{i}\right) & \frac{1}{q}\left(\delta_{ij-3}-\frac{\left(p_{i}-qA_{i}\right)\left(p_{j-3}-qA_{j-3}\right)}{c^{2}\left(\gamma m\right)^{2}}\right)\\
\hline \begin{array}{c}
\frac{\partial^{2}A_{k}}{\partial r^{j}\partial r_{i-3}}{\displaystyle \left(p_{k}-qA_{k}\right)}-\frac{\partial A_{k}}{\partial r_{i-3}}\frac{\partial A_{k}}{\partial r_{j}}\\
+\frac{\partial A_{k}}{\partial r_{i-3}}\left(p_{k}-qA_{k}\right)\frac{\partial A_{l}}{\partial r_{j}}{\displaystyle \frac{p_{l}-qA_{l}}{c^{2}\left(\gamma m\right)^{2}}}-\gamma m\frac{\partial^{2}\Phi}{\partial r_{i-3}\partial r_{j}}
\end{array} & \frac{\partial A_{j-3}}{\partial r_{i-3}}-\frac{\partial A_{k}}{\partial r_{i-3}}\frac{p_{k}-qA_{k}}{c^{2}\left(\gamma m\right)^{2}}\left(p_{j}-qA_{j}\right)
\end{array}\right)_{\boldsymbol{x}\left(t\right)}Y_{j}\,.
\end{equation}
The latter considerably simplifies if we consider optical elements,
where the kinetic energy of the particles is conserved (i.e., all
purely magnetic elements)
\begin{equation}
\frac{dY_{i}}{dt}=\frac{q}{\gamma m}\left(\begin{array}{c|c}
-\frac{\partial A_{i}}{\partial r_{j}} & \frac{1}{q}\delta_{ij-3}\\
\hline \frac{\partial^{2}A_{k}}{\partial r_{i-3}\partial r_{j}}\left(p_{k}-qA_{k}\right)-q\frac{\partial A_{k}}{\partial r_{i-3}}\frac{\partial A_{k}}{\partial r_{j}}-\gamma m\frac{\partial^{2}\Phi}{\partial r_{i-3}\partial r_{j}} & \frac{\partial A_{j-3}}{\partial r_{i-3}}
\end{array}\right)_{\boldsymbol{x}\left(t\right)}Y_{j}\,.\label{eq:rel_magn_flow}
\end{equation}
\end{widetext}Accordingly, we only have to replace $m\rightarrow\gamma m$
in the non-relativistic second order paraxial equations of motion
in this case
\begin{equation}
\gamma m\frac{d^{2}s_{i}}{dt^{2}}=-qF_{ij}\frac{ds_{j}}{dt}-q\frac{\partial F_{ij}}{\partial r_{j}}\frac{dr_{k}}{dt}s_{j}\,.
\end{equation}
The simplified relativistic JVE covers a wide range of CPO devices
since electric components are seldom employed at relativistic energies,
as their deflection efficiency falls with the beam energy.

In the Appendix we will describe a set of non-relativistic or purely
magnetic examples of elementary CPO devices using the previously developed
framework: the magnetic quadrupole, the magnetic round lens, the Wien
filter and the magnetic sector magnet. Here, we put the main focus
on illustrating the several computation steps, notably including gauge,
Magnus expansion and curved systems.

\section{Aberration Theory}

The iterative computation of higher order aberrations begins with
collecting quadratic terms ($n=2$) of $\varepsilon$ in (\ref{eq:perturb}),
corresponding to the first order aberration level. One obtains the
following coupled system of inhomogeneous first order differential
equations 
\begin{equation}
\frac{d\left(Y_{i}^{\left(2\right)}\right)}{dt}=\left(\frac{\partial X_{i}}{\partial x_{j}}\right)_{\boldsymbol{x}\left(t\right)}Y_{j}^{\left(2\right)}+\frac{1}{2}\left(\frac{\partial^{2}X_{i}}{\partial x_{j_{1}}\partial x_{j_{2}}}\right)_{\boldsymbol{x}\left(t\right)}Y_{j_{1}}^{\left(1\right)}Y_{j_{2}}^{\left(1\right)}\,,\label{eq:JVE2nd}
\end{equation}
which define the first order aberrations (see below). Noting that $Y_{i}^{\left(2\right)}\left(0\right)=0$  (i.e., starting conditions have been absorbed by paraxial trajectory)
the solution of such a first order matrix differential equation can
be formally written as \begin{widetext}
\begin{align}
Y_{i}^{\left(2\right)}\left(t\right) & =\frac{1}{2}\mathcal{M}_{ij}\left(t\right)\int_{0}^{t}\mathcal{M}_{jk}^{-1}\left(t_{1}\right)\left(\frac{\partial^{2}X_{k}}{\partial x_{j_{1}}\partial x_{j_{2}}}\right)_{\boldsymbol{x}\left(t_{1}\right)}Y_{j_{1}}^{\left(1\right)}\left(t_{1}\right)Y_{j_{2}}^{\left(1\right)}\left(t_{1}\right)\dd t_{1}\label{eq:2nd_order_M}\\
 & =\underset{\mathcal{M}_{il_{1}l_{2}}^{\left(2\right)}}{\underbrace{\left(\frac{1}{2}\mathcal{M}_{ij}\left(t\right)\int_{0}^{t}\left(J^{T}\mathcal{M}^{T}J\right)_{jk}\left(t_{1}\right)\left(\frac{\partial^{2}X_{k}}{\partial x_{j_{1}}\partial x_{j_{2}}}\right)_{\boldsymbol{x}\left(t_{1}\right)}\mathcal{M}_{j_{1}l_{1}}\left(t_{1}\right)\mathcal{M}_{j_{2}l_{2}}\left(t_{1}\right)\dd t_{1}\right)}}Y_{l_{1}}\left(0\right)Y_{l_{2}}\left(0\right)\,,\nonumber 
\end{align}
\end{widetext}where $\mathcal{M}$ denotes the paraxial transfer
matrices (\ref{eq:Transfer_Matrix}) as usual. On the second line, we exploited the inversion formula for symplectic matrices noted previously. The form of the solution
reveals the second order polynomial dependency of the second order
path deviation from the the initial path phase space coordinates.
Accordingly, we denote the second order ``transfer matrix'' (indeed
a vector-valued 2-form) by $\mathcal{M}^{\left(2\right)}$. 

Inspecting the definition of $\mathcal{M}^{\left(2\right)}$, we observe
that it is symmetric in the last two indices. Another set of fundamental
symmetries follows from the Jacobian of the Hamiltonian flow being
a Hamiltonian matrix and the Gaussian transfer matrices being symplectic,
ultimately leading to interdependencies between aberrations. Additional
restrictions may hold if the optical system has certain symmetries,
which restrict the non-zero entries of the derivatives of the Jacobian
of the Hamiltonian flow. Abstractly, they may be also derived from
symmetry conditions $\mathcal{M}^{\left(2\right)}=S\mathcal{M}^{\left(2\right)}S^{-1}S^{-1},$where
the matrices $S$ are representations of some symmetry operation and
operate on one particular index of $\mathcal{M}$ in the given order.
Another set of restrictions is imposed, when considering particular
pairs of planes, where the paraxial transfer matrices assume a particular
simple shape. E.g., in case of stigmatic imaging 
\begin{equation}
\mathcal{M}\left(t\right)=\left(\begin{array}{c|c}
A & 0\\
\hline B & A^{-1}
\end{array}\right)
\end{equation}
with $3\times3$ block matrices $A$ and $B$. To bring the above
integral to an analytically more tractable form one may finally apply
further approximations, such as the first order Magnus approximation,
i.e., $\mathcal{M}\approx\exp\left(\int\chi dt\right)$, or the weak
field approximation, i.e., $\mathcal{M}\approx1+\int\chi dt$, if
appropriate.

Further insight into the properties of (second order) path deviations
can be obtained by considering phase space areas spanned by deviation
vectors. To see that we first note that the directional derivative
along the Hamiltonian flow of the phase space areas spanned by $\boldsymbol{Y}^{\left(1\right)}+\boldsymbol{Y}^{\left(2\right)}$
and a Gaussian reference trajectory $\boldsymbol{Z}^{\left(1\right)}$
reads (see Appendix\ref{sec:Lie-Derivative})
\begin{equation}
\boldsymbol{X}\left\{ \omega\left(\boldsymbol{Y}^{\left(1\right)}+\boldsymbol{Y}^{\left(2\right)},\boldsymbol{Z}^{\left(1\right)}\right)\right\} =\omega\left(\mathcal{L}_{\boldsymbol{X}}\boldsymbol{Y}^{\left(2\right)},\boldsymbol{Z}^{\left(1\right)}\right)
\end{equation}
Inserting (\ref{eq:JVE2nd}) and pulling back to the design trajectory
parameterization $t$ then gives\begin{widetext}
\begin{equation}
\dv{t}\left[\omega\left(\boldsymbol{Y}^{\left(1\right)}+\boldsymbol{Y}^{\left(2\right)},\boldsymbol{Z}^{\left(1\right)}\right)\right]=\frac{1}{2}\omega\left(\left(\frac{\partial^{2}\boldsymbol{X}}{\partial x_{j_{1}}\partial x_{j_{2}}}\right)_{\boldsymbol{x}\left(t\right)}Y_{j_{1}}^{\left(1\right)}Y_{j_{2}}^{\left(1\right)},\boldsymbol{Z}^{\left(1\right)}\right)\,.
\end{equation}
This differential equation may be directly integrated taking into
account $\boldsymbol{Y}^{\left(2\right)}\left(t_{i}\right)=0$ yielding
\begin{equation}
\omega\left(\boldsymbol{Y}^{\left(1\right)}+\boldsymbol{Y}^{\left(2\right)},\boldsymbol{Z}^{\left(1\right)},t\right)-\omega\left(\boldsymbol{Y}^{\left(1\right)},\boldsymbol{Z}^{\left(1\right)},0\right)=\omega\left(\boldsymbol{Y}^{\left(2\right)},\boldsymbol{Z}^{\left(1\right)},t\right)=\frac{1}{2}\int_{0}^{t}\omega\left(\left(\frac{\partial^{2}\boldsymbol{X}}{\partial x_{j_{1}}\partial x_{j_{2}}}\right)_{\boldsymbol{x}\left(t\right)}Y_{j_{1}}^{\left(1\right)}Y\left(t_{1}\right)_{j_{2}}^{\left(1\right)}\left(t_{1}\right),\boldsymbol{Z}^{\left(1\right)}\left(t_{1}\right)\right)\dd t_{1}\label{eq:P2form2ndorder}
\end{equation}
\end{widetext}These relations replace the Lagrange invariants of
the paraxial case, based on $\dv{t}\left[\omega\left(\boldsymbol{Y}^{\left(1\right)},\boldsymbol{Z}^{\left(1\right)}\right)\right]=0$.
Indeed, (\ref{eq:P2form2ndorder}) correspond to the expressions evaluated
in the so-called eikonal methods (e.g., \citep{Hawkes(1996)}). The
latter essentially solve (\ref{eq:P2form2ndorder}) for maximally
six (often fewer are sufficient) linearly independent referenze trajectories
$\boldsymbol{Z}_{1..6}^{\left(1\right)}$, which allows to determine
all six components of $\boldsymbol{Y}^{\left(2\right)}$by solving
the 6 equations $\omega\left(\boldsymbol{Y}^{\left(2\right)},\boldsymbol{Z}_{1..6}^{\left(1\right)},t\right)$
for $\boldsymbol{Y}^{\left(2\right)}$. The latter is always possible
as the two-form $\omega$ is non-degenerate. Note, furthermore, that
the Gaussian reference trajectories $\boldsymbol{Z}^{\left(1\right)}$
may be in principle chosen such to coincide with the second order
trajectory $\boldsymbol{Y}^{\left(1\right)}+\boldsymbol{Y}^{\left(2\right)}$
in only one phase space coordinate (either position or momentum) in
the initial plane and another one in the final plane (or any other,
such a the aperture plane), respectively. The evaluation of (\ref{eq:P2form2ndorder})
using such mixed boundary conditions is, however, not straight forward
as the value of $\boldsymbol{Y}^{\left(2\right)}$ at the second plane
is only implicitely defined.

We finally have to consider two important aspects of the above aberration
theory: (A) Additional set of deviations introduced by the longitudinal
position deviations with respect to some predefined final plane (e.g.,
image plane). The longitudinal deviations correspond to a time lag
for the particle to reach the final plane which can lead to additional
aberrations after propagation to that plane. (B) The canonical momentum
deviation considered above needs to be transformed into kinetic momentum
to obtain experimentally observable quantities. We simplify both considerations
by discussing initial and final planes, which are outside of the fields
of the optical elements (constant non-zero potentials remain possible).
That is a minor restriction considering that we are mostly interested
into the properties of optical elements with respect to planes lying
outside. Notable exceptions are, for instance, the asymptotic properties
of objective lenses in a transmission electron microscope, where the
object plane is immersed in the magnetic field. Under these prerequisites
the trajectories in the initial and final plane are straight and kinetic
and canonical momentum are equivalent. Consequently, we only have
to add a first order correction due to transit time differences $\delta t$
to some predefined final plane, reading
\begin{align}
Y_{i}{}^{\left(2t\right)}\left(t\right) & \approx-\chi_{ij}\left(t\right)Y_{j}^{\left(1\right)}\left(t\right)\delta t\label{eq:2ndorder_timelag}\\
 & =-\frac{1}{mv}\left(\begin{array}{c|c}
0 & I_{3}\\
\hline 0 & 0
\end{array}\right)_{ij}Y_{j}^{\left(1\right)}\left(t\right)Y_{3}^{\left(1\right)}\left(t\right)\nonumber \\
 & =\underset{\mathcal{M}_{il_{1}l_{2}}^{\left(2t\right)}}{\underbrace{-\frac{1}{mv}\left(\begin{array}{c|c}
0 & I_{3}\\
\hline 0 & 0
\end{array}\right)_{ij}\mathcal{M}_{jl_{1}}\mathcal{M}_{3l_{2}}}}Y_{l_{1}}\left(0\right)Y_{l_{2}}\left(0\right)\,,\nonumber 
\end{align}
where the explicit form of the free space Jacobian of the Hamiltonian
flow has been inserted. It is furthermore useful to symmetrize $\mathcal{M}^{\left(2t\right)}$ in the last two indices as there is no preferred choice, i.e., $\mathcal{M}_{il_{1}l_{2}}^{\left(2t\right)}=\nicefrac{\left(\mathcal{M}_{il_{1}l_{2}}^{\left(2t\right)}+\mathcal{M}_{il_{2}l_{1}}^{\left(2t\right)}\right)}{2}$. In order to obtain aberration coefficients,
which we define here as the dependency of path deviations of a certain
order in some final plane on homogeneous polynomials of corresponding
order (here 2) of position and kinetic momentum in some initial plane,
we finally have to add the above two contributions, i.e., $C_{ijj}^{\left(2\right)}=M_{ijj}^{\left(2\right)}+M_{ijj}^{\left(2t\right)}$
and $C_{ijk\neq j}^{\left(2\right)}=2\left(M_{ijk}^{\left(2\right)}+M_{ijk}^{\left(2t\right)}\right)$,
where the factor 2 accounts for the combinatorial multiplicity of
the two symmetric terms in $M^{\left(2\right)}$.

The above sketched program may now be extended to higher order aberrations
with little modifications: The inhomogeneous system of first order
differential equations defining the third order path deviations in
phase space read\begin{widetext}
\begin{equation}
\dv{Y_{i}^{\left(3\right)}}{t}\approx\left(\frac{\partial X_{i}}{\partial x_{j}}\right)_{\boldsymbol{x}\left(t\right)}Y_{j}^{\left(3\right)}+\left(\frac{\partial^{2}X_{i}}{\partial x_{j_{1}}\partial x_{j_{2}}}\right)_{\boldsymbol{x}\left(t\right)}Y_{j_{1}}^{\left(1\right)}Y_{j_{2}}^{\left(2\right)}+\frac{1}{6}\left(\frac{\partial^{3}X_{i}}{\partial x_{j_{1}}\partial x_{j_{2}}\partial x_{j_{3}}}\right)_{\boldsymbol{x}\left(t\right)}Y_{j_{1}}^{\left(1\right)}Y_{j_{2}}^{\left(1\right)}Y_{j_{3}}^{\left(1\right)}\,.
\end{equation}
Again solutions can be obtained by direct integration
\begin{align}
Y_{i}^{\left(3\right)}\left(t\right) & =\mathcal{M}_{ij}\left(t\right)\int_{0}^{t}\left(J^{T}\mathcal{M}^{T}J\right)_{jk}\left(t_{1}\right)\left(\left(\frac{\partial^{2}X_{k}}{\partial x_{j_{1}}\partial x_{j_{2}}}\right)_{\boldsymbol{x}\left(t_{1}\right)}Y_{j_{1}}^{\left(1\right)}\left(t_{1}\right)Y_{j_{2}}^{\left(2\right)}\left(t_{1}\right)\right.\label{eq:3rd_order_M}\\
 & +\left.\frac{1}{6}\left(\frac{\partial^{2}X_{k}}{\partial x_{j_{1}}\partial x_{j_{2}}x_{j_{1}}\partial x_{j_{3}}}\right)_{\boldsymbol{x}\left(t_{1}\right)}Y_{j_{1}}^{\left(1\right)}\left(t_{1}\right)Y_{j_{2}}^{\left(1\right)}\left(t_{1}\right)Y_{j_{3}}^{\left(1\right)}\left(t_{1}\right)\right)\dd t_{1}\nonumber \\
 & =\left(\mathcal{M}_{ij}\left(t\right)\int_{0}^{t}\left(J^{T}\mathcal{M}^{T}J\right)_{jk}\left(t_{1}\right)\left(\frac{\partial^{2}X_{k}}{\partial x_{j_{1}}\partial x_{j_{2}}}\right)_{\boldsymbol{x}\left(t_{1}\right)}\mathcal{M}_{j_{1}l_{1}}\left(t_{1}\right)\mathcal{M}_{j_{2}l_{2}l_{3}}^{\left(2\right)}\left(t_{1}\right)\dd t_{1}\right.\nonumber \\
 & +\left.\frac{1}{6}\mathcal{M}_{ij}\left(t\right)\int_{0}^{t}\left(J^{T}\mathcal{M}^{T}J\right)_{jk}\left(t_{1}\right)\left(\frac{\partial^{2}X_{k}}{\partial x_{j_{1}}\partial x_{j_{2}}x_{j_{1}}\partial x_{j_{3}}}\right)_{\boldsymbol{x}\left(t_{1}\right)}\mathcal{M}_{j_{1}l_{1}}\left(t_{1}\right)\mathcal{M}_{j_{2}l_{2}}\left(t_{1}\right)\mathcal{M}_{j_{3}l_{3}}\left(t_{1}\right)\dd t_{1}\right)Y_{l_{1}}^{\left(1\right)}\left(0\right)Y_{l_{2}}^{\left(1\right)}\left(0\right)Y_{l_{3}}^{\left(1\right)}\left(0\right)\,.\nonumber 
\end{align}
\end{widetext}The expression on the third line may be further symmetrized
with respect to $l_{1...3}$ as there is no preferred choice in the
ordering of the corresponding $\boldsymbol{Y}_{l_{1...3}}\left(0\right)$.
Again, the whole prefactor of the third order polynomial of initial
beam parameters can be referred to as generalized transfer matrix
$\mathcal{M}^{\left(3\right)}$. Additional corrections due to transit
time lags (taking into account that derivatives of $\chi$ in vacuum
are zero) read 
\begin{align}
\boldsymbol{Y}^{\left(3t\right)}\left(t\right) & \approx-\frac{1}{m}\left(\begin{array}{c|c}
0 & I_{3}\\
\hline 0 & 0
\end{array}\right)\boldsymbol{Y}^{\left(1\right)}\left(t\right)\delta t^{\left(2\right)}\label{eq:3rdordertime}\\
 & =-\frac{1}{mv}\left(\begin{array}{c|c}
0 & I_{3}\\
\hline 0 & 0
\end{array}\right)_{ij}Y_{j}^{\left(1\right)}\left(t\right)Y_{3}^{\left(2\right)}\left(t\right)\nonumber \\
 & =\underset{\mathcal{M}_{il_{1}l_{2}l_{3}}^{\left(3t\right)}}{\underbrace{-\frac{1}{mv}\left(\begin{array}{c|c}
0 & I_{3}\\
\hline 0 & 0
\end{array}\right)_{ij}\mathcal{M}_{jl_{1}}\mathcal{M}_{3l_{2}l_{3}}^{\left(2\right)}}}Y_{l_{1}}^{\left(1\right)}\left(0\right)Y_{l_{2}}^{\left(1\right)}\left(0\right)Y_{l_{3}}^{\left(1\right)}\left(0\right)\,.\nonumber 
\end{align}
Again, the unsymmetrized expression is shown. Note, that a technical complication in the above expressions is given
by the presence of second order deviations $Y_{j_{2}}^{\left(2\right)}$
in the integral in the first line of (\ref{eq:3rd_order_M}) and the
second line of (\ref{eq:3rdordertime}). These terms are sometimes
referred to as intrinsic combination aberrations (as lower order combine
to higher order aberrations), and can be calculated with the help
of (\ref{eq:2nd_order_M}), leading, e.g., to the double integrals
on the third line of (\ref{eq:3rd_order_M}). The third and higher
aberration orders (i.e., $\boldsymbol{Y}^{\left(n>3\right)}$deviation
vectors) can be derived exactly along the same lines. In Appendix
\ref{sec:Examples} we will discuss the first order aberrations of
the quadrupole and Wien filter illustrating the above framework.

\section{Discussion and Outlook}

To sum up, an exhaustive yet compact description of charged particle
optics rooted in Hamiltonian mechanics has been presented. The main
features are incorporation of transversal and longitudinal (i.e.,
chromatic) path deviations in static and time-dependent fields on
the same footing, a unified and simple description of both straight
and curved systems, the straightforward computation of transfer matrices
up to any perturbation order (i.e., aberrations), and an efficient
geometric integration scheme. The framework shall be useful for characterization
of general CPO devices with regard to two aspects: (A) Employing closed
analytical approximations for the paraxial transfer matrices $M$
derived from the Magnus expansion allows to write down analytical
expressions for the aberration coefficients. Notwithstanding their
approximate character such expressions are useful to quickly evaluate
the importance of certain aberrations and possible reduction strategies.
(B) Employing ``exact'' paraxial transfer matrices (either obtained
by multi-step methods or geometric integrators) the aberration integrals
may be efficiently integrated numerically, yielding fast and accurate
perturbative trajectory solvers. 

\section{Acknowledgement}

We are deeply indebted to Stephan Uhlemann, Heiko Müller and Peter
Hawkes for invaluable discussions and comments. We furthermore acknowledge
funding from the European Research Council (ERC) under the Horizon
2020 research and innovation program of the European Union (grant
agreement no. 715620).

\section{Bibliogaphy}

\bibliographystyle{apsrev4-2}
\bibliography{Lie}

%apsrev4-2.bst 2019-01-14 (MD) hand-edited version of apsrev4-1.bst
%Control: key (0)
%Control: author (72) initials jnrlst
%Control: editor formatted (1) identically to author
%Control: production of article title (-1) disabled
%Control: page (0) single
%Control: year (1) truncated
%Control: production of eprint (0) enabled
\begin{thebibliography}{25}%
\makeatletter
\providecommand \@ifxundefined [1]{%
 \@ifx{#1\undefined}
}%
\providecommand \@ifnum [1]{%
 \ifnum #1\expandafter \@firstoftwo
 \else \expandafter \@secondoftwo
 \fi
}%
\providecommand \@ifx [1]{%
 \ifx #1\expandafter \@firstoftwo
 \else \expandafter \@secondoftwo
 \fi
}%
\providecommand \natexlab [1]{#1}%
\providecommand \enquote  [1]{``#1''}%
\providecommand \bibnamefont  [1]{#1}%
\providecommand \bibfnamefont [1]{#1}%
\providecommand \citenamefont [1]{#1}%
\providecommand \href@noop [0]{\@secondoftwo}%
\providecommand \href [0]{\begingroup \@sanitize@url \@href}%
\providecommand \@href[1]{\@@startlink{#1}\@@href}%
\providecommand \@@href[1]{\endgroup#1\@@endlink}%
\providecommand \@sanitize@url [0]{\catcode `\\12\catcode `\$12\catcode
  `\&12\catcode `\#12\catcode `\^12\catcode `\_12\catcode `\%12\relax}%
\providecommand \@@startlink[1]{}%
\providecommand \@@endlink[0]{}%
\providecommand \url  [0]{\begingroup\@sanitize@url \@url }%
\providecommand \@url [1]{\endgroup\@href {#1}{\urlprefix }}%
\providecommand \urlprefix  [0]{URL }%
\providecommand \Eprint [0]{\href }%
\providecommand \doibase [0]{https://doi.org/}%
\providecommand \selectlanguage [0]{\@gobble}%
\providecommand \bibinfo  [0]{\@secondoftwo}%
\providecommand \bibfield  [0]{\@secondoftwo}%
\providecommand \translation [1]{[#1]}%
\providecommand \BibitemOpen [0]{}%
\providecommand \bibitemStop [0]{}%
\providecommand \bibitemNoStop [0]{.\EOS\space}%
\providecommand \EOS [0]{\spacefactor3000\relax}%
\providecommand \BibitemShut  [1]{\csname bibitem#1\endcsname}%
\let\auto@bib@innerbib\@empty
%</preamble>
\bibitem [{\citenamefont {Scherzer}(1933)}]{Scherzer1933}%
  \BibitemOpen
  \bibfield  {author} {\bibinfo {author} {\bibfnamefont {O.}~\bibnamefont
  {Scherzer}},\ }\href {https://doi.org/https://doi.org/10.1007/BF02055909}
  {\bibfield  {journal} {\bibinfo  {journal} {{Zeitschrift f\"ur Physik}}\
  }\textbf {\bibinfo {volume} {80}},\ \bibinfo {pages} {193} (\bibinfo {year}
  {1933})}\BibitemShut {NoStop}%
\bibitem [{\citenamefont {Glaser}(1933)}]{Glaser1933}%
  \BibitemOpen
  \bibfield  {author} {\bibinfo {author} {\bibfnamefont {W.}~\bibnamefont
  {Glaser}},\ }\href {https://doi.org/10.1007/BF02057307} {\bibfield  {journal}
  {\bibinfo  {journal} {{Zeitschrift f\"ur Physik}}\ }\textbf {\bibinfo
  {volume} {80}},\ \bibinfo {pages} {451} (\bibinfo {year} {1933})}\BibitemShut
  {NoStop}%
\bibitem [{\citenamefont {Sturrock}\ and\ \citenamefont
  {Allibone}(1951)}]{Sturrock1951}%
  \BibitemOpen
  \bibfield  {author} {\bibinfo {author} {\bibfnamefont {P.~A.}\ \bibnamefont
  {Sturrock}}\ and\ \bibinfo {author} {\bibfnamefont {T.~E.}\ \bibnamefont
  {Allibone}},\ }\href
  {https://royalsocietypublishing.org/doi/abs/10.1098/rspa.1951.0245}
  {\bibfield  {journal} {\bibinfo  {journal} {Proceedings of the Royal Society
  of London. Series A. Mathematical and Physical Sciences}\ }\textbf {\bibinfo
  {volume} {210}},\ \bibinfo {pages} {269} (\bibinfo {year}
  {1951})}\BibitemShut {NoStop}%
\bibitem [{\citenamefont {Hawkes}\ and\ \citenamefont
  {Kasper}(1996)}]{Hawkes(1996)}%
  \BibitemOpen
  \bibfield  {author} {\bibinfo {author} {\bibfnamefont {P.~W.}\ \bibnamefont
  {Hawkes}}\ and\ \bibinfo {author} {\bibfnamefont {E.}~\bibnamefont
  {Kasper}},\ }\href@noop {} {\emph {\bibinfo {title} {{Principles of Electron
  Optics Vol. 1: Basic geometrical optics}}}},\ \bibinfo {series} {Principles
  of Electron Optics}\ No.\ \bibinfo {number} {Bd. 3}\ (\bibinfo  {publisher}
  {Academic Press},\ \bibinfo {year} {1996})\BibitemShut {NoStop}%
\bibitem [{\citenamefont {Rose}(2004)}]{Rose2004}%
  \BibitemOpen
  \bibfield  {author} {\bibinfo {author} {\bibfnamefont {H.}~\bibnamefont
  {Rose}},\ }\href {https://doi.org/https://doi.org/10.1016/j.nima.2003.11.115}
  {\bibfield  {journal} {\bibinfo  {journal} {Nuclear Instruments and Methods
  in Physics Research Section A: Accelerators, Spectrometers, Detectors and
  Associated Equipment}\ }\textbf {\bibinfo {volume} {519}},\ \bibinfo {pages}
  {12} (\bibinfo {year} {2004})}\BibitemShut {NoStop}%
\bibitem [{\citenamefont {Dragt}(1982)}]{Dragt1982}%
  \BibitemOpen
  \bibfield  {author} {\bibinfo {author} {\bibfnamefont {A.~J.}\ \bibnamefont
  {Dragt}},\ }\href@noop {} {\bibfield  {journal} {\bibinfo  {journal} {AIP
  Conference Proceedings}\ }\textbf {\bibinfo {volume} {87}},\ \bibinfo {pages}
  {147} (\bibinfo {year} {1982})}\BibitemShut {NoStop}%
\bibitem [{\citenamefont {Dragt}(1987)}]{Dragt1987}%
  \BibitemOpen
  \bibfield  {author} {\bibinfo {author} {\bibfnamefont {A.~J.}\ \bibnamefont
  {Dragt}},\ }\href
  {https://doi.org/https://doi.org/10.1016/0168-9002(87)90916-8} {\bibfield
  {journal} {\bibinfo  {journal} {Nuclear Instruments and Methods in Physics
  Research Section A: Accelerators, Spectrometers, Detectors and Associated
  Equipment}\ }\textbf {\bibinfo {volume} {258}},\ \bibinfo {pages} {339}
  (\bibinfo {year} {1987})}\BibitemShut {NoStop}%
\bibitem [{\citenamefont {Dragt}\ \emph {et~al.}(1988)\citenamefont {Dragt},
  \citenamefont {Neri}, \citenamefont {Rangarajan}, \citenamefont {Douglas},
  \citenamefont {Healy},\ and\ \citenamefont {Ryne}}]{Dragt1988}%
  \BibitemOpen
  \bibfield  {author} {\bibinfo {author} {\bibfnamefont {A.~J.}\ \bibnamefont
  {Dragt}}, \bibinfo {author} {\bibfnamefont {F.}~\bibnamefont {Neri}},
  \bibinfo {author} {\bibfnamefont {G.}~\bibnamefont {Rangarajan}}, \bibinfo
  {author} {\bibfnamefont {D.~R.}\ \bibnamefont {Douglas}}, \bibinfo {author}
  {\bibfnamefont {L.~M.}\ \bibnamefont {Healy}},\ and\ \bibinfo {author}
  {\bibfnamefont {R.~D.}\ \bibnamefont {Ryne}},\ }\href
  {https://www.annualreviews.org/doi/pdf/10.1146/annurev.ns.38.120188.002323}
  {\bibfield  {journal} {\bibinfo  {journal} {Annual Review of Nuclear and
  Particle Science}\ }\textbf {\bibinfo {volume} {38}},\ \bibinfo {pages} {455}
  (\bibinfo {year} {1988})}\BibitemShut {NoStop}%
\bibitem [{\citenamefont {Cary}(1977)}]{Cary1977}%
  \BibitemOpen
  \bibfield  {author} {\bibinfo {author} {\bibfnamefont {J.~R.}\ \bibnamefont
  {Cary}},\ }\href@noop {} {\bibfield  {journal} {\bibinfo  {journal} {Journal
  of Mathematical Physics}\ }\textbf {\bibinfo {volume} {18}},\ \bibinfo
  {pages} {2432} (\bibinfo {year} {1977})}\BibitemShut {NoStop}%
\bibitem [{\citenamefont {Cary}(1981)}]{Cary1981}%
  \BibitemOpen
  \bibfield  {author} {\bibinfo {author} {\bibfnamefont {J.~R.}\ \bibnamefont
  {Cary}},\ }\href@noop {} {\bibfield  {journal} {\bibinfo  {journal} {Physics
  Reports}\ }\textbf {\bibinfo {volume} {79}},\ \bibinfo {pages} {129}
  (\bibinfo {year} {1981})}\BibitemShut {NoStop}%
\bibitem [{\citenamefont {Radli\v{c}ka}(2009)}]{Radlicka2009}%
  \BibitemOpen
  \bibfield  {author} {\bibinfo {author} {\bibfnamefont {T.}~\bibnamefont
  {Radli\v{c}ka}}\ }(\bibinfo  {publisher} {Elsevier},\ \bibinfo {year}
  {2009})\ pp.\ \bibinfo {pages} {241--362}\BibitemShut {NoStop}%
\bibitem [{\citenamefont {Hawkes}\ and\ \citenamefont
  {Kasper}(1988)}]{Hawkes(1988)}%
  \BibitemOpen
  \bibfield  {author} {\bibinfo {author} {\bibfnamefont {P.~W.}\ \bibnamefont
  {Hawkes}}\ and\ \bibinfo {author} {\bibfnamefont {E.}~\bibnamefont
  {Kasper}},\ }\href@noop {} {\emph {\bibinfo {title} {{Principles of Electron
  Optics Vol.2: Applied Geometrical Optics}}}}\ (\bibinfo  {publisher}
  {Elsevier Science},\ \bibinfo {year} {1988})\BibitemShut {NoStop}%
\bibitem [{\citenamefont {Frankel}(2004)}]{Frankel2004}%
  \BibitemOpen
  \bibfield  {author} {\bibinfo {author} {\bibfnamefont {T.}~\bibnamefont
  {Frankel}},\ }\href {https://books.google.de/books?id=DUnjs6nEn8wC} {\emph
  {\bibinfo {title} {{The Geometry of Physics: An Introduction}}}}\ (\bibinfo
  {publisher} {Cambridge University Press},\ \bibinfo {year}
  {2004})\BibitemShut {NoStop}%
\bibitem [{\citenamefont {Dragt}\ and\ \citenamefont {Finn}(1976)}]{Dragt1976}%
  \BibitemOpen
  \bibfield  {author} {\bibinfo {author} {\bibfnamefont {A.~J.}\ \bibnamefont
  {Dragt}}\ and\ \bibinfo {author} {\bibfnamefont {J.~M.}\ \bibnamefont
  {Finn}},\ }\href@noop {} {\bibfield  {journal} {\bibinfo  {journal} {Journal
  of Mathematical Physics}\ }\textbf {\bibinfo {volume} {17}},\ \bibinfo
  {pages} {2215} (\bibinfo {year} {1976})}\BibitemShut {NoStop}%
\bibitem [{\citenamefont {Magnus}(1954)}]{Magnus1954}%
  \BibitemOpen
  \bibfield  {author} {\bibinfo {author} {\bibfnamefont {W.}~\bibnamefont
  {Magnus}},\ }\href
  {https://onlinelibrary.wiley.com/doi/abs/10.1002/cpa.3160070404} {\bibfield
  {journal} {\bibinfo  {journal} {Communications on Pure and Applied
  Mathematics}\ }\textbf {\bibinfo {volume} {7}},\ \bibinfo {pages} {649}
  (\bibinfo {year} {1954})}\BibitemShut {NoStop}%
\bibitem [{\citenamefont {Leimkuhler}\ and\ \citenamefont
  {Reich}(2005)}]{Leimkuhler2005}%
  \BibitemOpen
  \bibfield  {author} {\bibinfo {author} {\bibfnamefont {B.}~\bibnamefont
  {Leimkuhler}}\ and\ \bibinfo {author} {\bibfnamefont {S.}~\bibnamefont
  {Reich}},\ }\href {https://doi.org/10.1017/CBO9780511614118} {\emph {\bibinfo
  {title} {{Simulating Hamiltonian Dynamics}}}},\ Cambridge Monographs on
  Applied and Computational Mathematics\ (\bibinfo  {publisher} {Cambridge
  University Press},\ \bibinfo {year} {2005})\BibitemShut {NoStop}%
\bibitem [{\citenamefont {Courant}\ \emph {et~al.}(1952)\citenamefont
  {Courant}, \citenamefont {Livingston},\ and\ \citenamefont
  {Snyder}}]{Courant1952}%
  \BibitemOpen
  \bibfield  {author} {\bibinfo {author} {\bibfnamefont {E.~D.}\ \bibnamefont
  {Courant}}, \bibinfo {author} {\bibfnamefont {M.~S.}\ \bibnamefont
  {Livingston}},\ and\ \bibinfo {author} {\bibfnamefont {H.~S.}\ \bibnamefont
  {Snyder}},\ }\href {https://doi.org/10.1103/PhysRev.88.1190} {\bibfield
  {journal} {\bibinfo  {journal} {Physical Review}\ }\textbf {\bibinfo {volume}
  {88}},\ \bibinfo {pages} {1190} (\bibinfo {year} {1952})}\BibitemShut
  {NoStop}%
\bibitem [{\citenamefont {Hawkes}(1970)}]{Hawkes1970}%
  \BibitemOpen
  \bibfield  {author} {\bibinfo {author} {\bibfnamefont {P.~W.}\ \bibnamefont
  {Hawkes}},\ }\href {https://books.google.de/books?id=4I0qAAAAYAAJ} {\emph
  {\bibinfo {title} {{Quadrupoles in Electron Lens Design}}}},\ Advances in
  electronics and electron physics: Supplement\ (\bibinfo  {publisher}
  {Academic Press},\ \bibinfo {year} {1970})\BibitemShut {NoStop}%
\bibitem [{\citenamefont {Hawkes}(1966)}]{Hawkes1966}%
  \BibitemOpen
  \bibfield  {author} {\bibinfo {author} {\bibfnamefont {P.~W.}\ \bibnamefont
  {Hawkes}},\ }\href {https://books.google.de/books?id=pP9PAQAAIAAJ} {\emph
  {\bibinfo {title} {{Quadrupole Optics}}}},\ Springer Tracts in Modern
  Physics\ (\bibinfo  {publisher} {Springer Berlin Heidelberg},\ \bibinfo
  {year} {1966})\BibitemShut {NoStop}%
\bibitem [{\citenamefont {Busch}(1926)}]{Busch(1926)}%
  \BibitemOpen
  \bibfield  {author} {\bibinfo {author} {\bibfnamefont {H.}~\bibnamefont
  {Busch}},\ }\href@noop {} {\bibfield  {journal} {\bibinfo  {journal} {Annalen
  der Physik}\ }\textbf {\bibinfo {volume} {81}},\ \bibinfo {pages} {974}
  (\bibinfo {year} {1926})}\BibitemShut {NoStop}%
\bibitem [{\citenamefont {Tsuno}\ and\ \citenamefont
  {Ioanoviciu}(2013)}]{Tsuno2013}%
  \BibitemOpen
  \bibfield  {author} {\bibinfo {author} {\bibfnamefont {K.}~\bibnamefont
  {Tsuno}}\ and\ \bibinfo {author} {\bibfnamefont {D.}~\bibnamefont
  {Ioanoviciu}},\ }\href@noop {} {\emph {\bibinfo {title} {The Wien Filter}}}\
  (\bibinfo {year} {2013})\BibitemShut {NoStop}%
\bibitem [{\citenamefont {Castaing}\ \emph {et~al.}(1967)\citenamefont
  {Castaing}, \citenamefont {Hennequin}, \citenamefont {Henry},\ and\
  \citenamefont {Slodzian}}]{Enge1967}%
  \BibitemOpen
  \bibfield  {author} {\bibinfo {author} {\bibfnamefont {R.}~\bibnamefont
  {Castaing}}, \bibinfo {author} {\bibfnamefont {J.~F.}\ \bibnamefont
  {Hennequin}}, \bibinfo {author} {\bibfnamefont {L.}~\bibnamefont {Henry}},\
  and\ \bibinfo {author} {\bibfnamefont {G.}~\bibnamefont {Slodzian}},\
  }\bibinfo {title} {{Focusing of Charged Particles}}\ (\bibinfo  {publisher}
  {Academic Press},\ \bibinfo {year} {1967})\ Chap.\ \bibinfo {chapter} {The
  Magnetic Prism as an Optical System}, p.\ \bibinfo {pages} {265}\BibitemShut
  {NoStop}%
\bibitem [{\citenamefont {Wollnik}\ and\ \citenamefont
  {Hermann}(1987)}]{Wollnik1987}%
  \BibitemOpen
  \bibfield  {author} {\bibinfo {author} {\bibfnamefont {H.}~\bibnamefont
  {Wollnik}}\ and\ \bibinfo {author} {\bibfnamefont {W.}~\bibnamefont
  {Hermann}},\ }\href {https://books.google.de/books?id=TxeZv3UkbpoC} {\emph
  {\bibinfo {title} {{Optics of Charged Particles}}}},\ Mathematics in Science
  and\ (\bibinfo  {publisher} {Academic Press},\ \bibinfo {year}
  {1987})\BibitemShut {NoStop}%
\bibitem [{\citenamefont {Enge}(1967)}]{Enge1967a}%
  \BibitemOpen
  \bibfield  {author} {\bibinfo {author} {\bibfnamefont {H.~A.}\ \bibnamefont
  {Enge}},\ }\bibinfo {title} {{Focusing of Charged Particles}}\ (\bibinfo
  {publisher} {Academic Press},\ \bibinfo {year} {1967})\ Chap.\ \bibinfo
  {chapter} {Deflecting Magnets}, p.\ \bibinfo {pages} {203}\BibitemShut
  {NoStop}%
\bibitem [{\citenamefont {Barber}(1933)}]{Barber1933}%
  \BibitemOpen
  \bibfield  {author} {\bibinfo {author} {\bibfnamefont {N.~F.}\ \bibnamefont
  {Barber}},\ }\href@noop {} {\bibfield  {journal} {\bibinfo  {journal} {Proc.
  Leeds Philos. Lit. Soc. Sci. Sect.}\ }\textbf {\bibinfo {volume} {2}},\
  \bibinfo {pages} {427} (\bibinfo {year} {1933})}\BibitemShut {NoStop}%
\end{thebibliography}%

\appendix

\section{Lie Derivative\label{sec:Lie-Derivative}}

This section introduces the notion of the Lie derivative as employed
in the main text. The notation is adapted from \citep{Frankel2004},
which also contains a comprehensive introduction into the topic. The
Lie derivative of a vector field is defined as
\begin{equation}
{\displaystyle (\mathcal{L}_{\boldsymbol{X}}\boldsymbol{Y})_{\boldsymbol{x}}=\left.\frac{d}{dt}\right|_{t=0}\left((\varphi_{-t})_{*}\boldsymbol{Y}_{\varphi_{t}(\boldsymbol{x})}\right).}
\end{equation}
Here, the 1-parameter group (flow) $\varphi$ is given by
\begin{equation}
\frac{d}{dt}\varphi=\boldsymbol{X}
\end{equation}
and $\varphi{}_{*}$ denotes its differential. For our purposes $\boldsymbol{Y}$
needs to be defined along the flow generated by $\boldsymbol{X}$only
(which corresponds to the optical axis). The coordinate expression
for the Lie derivative reads
\begin{align}
\left(\mathcal{L}_{\boldsymbol{X}}\boldsymbol{Y}\right)_{i} & =\frac{\partial Y^{i}}{\partial x^{j}}X^{j}-\frac{\partial X_{i}}{\partial x_{j}}Y_{j}\\
 & =\frac{dY_{i}}{dt}-\frac{\partial X_{i}}{\partial x_{j}}Y_{j}\,.\nonumber 
\end{align}
In the main text we are concerned with so-called invariant vector
fields fulfilling $\mathcal{L}_{\boldsymbol{X}}\boldsymbol{Y}=0$,
for which the Lie derivative of a $p$-dimensional form evaluated
on invariant vector fields $\boldsymbol{Y}$reads 
\begin{equation}
\mathcal{L}_{\boldsymbol{X}}\omega^{p}\left(\boldsymbol{Y}_{1},\ldots,\boldsymbol{Y}_{p}\right)=\frac{d}{dt}\left[\omega_{\varphi_{t}x}^{p}\left(\boldsymbol{Y}_{1},\ldots,\boldsymbol{Y}_{p}\right)\right]\,.
\end{equation}
Consequently, in our phase space setting, $\mathcal{L}_{\boldsymbol{X}}\omega\left(\boldsymbol{Y}_{1},\boldsymbol{Y}_{2}\right)$
measures the derivative, with respect to motion along the optical
axis, of phase space areas spanned by vector fields $\boldsymbol{Y}$,
which equates to zero according to Liouville's theorem. Finally, from
the ``Leibniz'' rule for the Lie derivative we can equate the directional
derivative of the Poincare 2-form of arbitrary vector fields along
the optical axis with Poincare 2-forms of Lie derivatives\citep{Frankel2004}
\begin{equation}
\boldsymbol{X}\left\{ \omega\left(\boldsymbol{Y},\boldsymbol{Z}\right)\right\} =\omega\left(\mathcal{L}_{\boldsymbol{X}}\boldsymbol{Y},\boldsymbol{Z}\right)+\omega\left(\boldsymbol{Y},\mathcal{L}_{\boldsymbol{X}}\boldsymbol{Z}\right)\,.
\end{equation}

\section{Examples\label{sec:Examples}}

In this Appendix we illustrate the various working principles of the
above framework at the example of four basic CPO elements, which are
important in numerous applications: (A) the magnetic quadrupole, (B)
the round magnetic lens, (C) the Wien filter, and (D) the sector magnet.
The first three elements have a straight optical axis, whereas the
sector magnet has a curved optical axis. The first two examples show
no variation in the transverse trajectory coordinates with respect
to an energy change in the paraxial limit. The last two shows a first
order dispersion, because of which they are used in energy filters
and monochromator applications. In the specific case of electron optics,
the first, second and fourth are purely magnetic and typically operate
with relativistic electrons, whereas the Wien filter typically resides,
where the electrons are rather slow (non-relativistic). As the pure
relativistic magnetic systems can be described by the simplified relativistic
Hamiltonian flow (\ref{eq:rel_magn_flow}), we implicitly assume relativistic
masses and omit $\gamma$ in the following expressions.

The level of the discussion will vary from element to element to keep
the length at bay while illustrating all important aspects. E.g.,
quadrupole and Wien filter contain evaluation of aberrations, which
are omitted for the lens and sector magnet. The lens on the other
hand is the only example in which the field varies with the optical
axis. For the other three cases we will only consider homogeneous
fields, which are sharply cut-off at the entrance and exit plane of
the respective device (such a neglection of fringing fields does not
adequately describes real elements, we will furthermore not elaborate
on the implications of the sharp cut-off on aberrations). The sector
magnet section mainly focusses on the implications of a curved optical
axis. We generally keep the discussion of the final results short
as they are typically well-known and our main goal is the illustration
of the machinery developed in the manuscript.

\subsection{Magnetic Quadrupole}

\begin{figure}
\includegraphics[width=1\columnwidth]{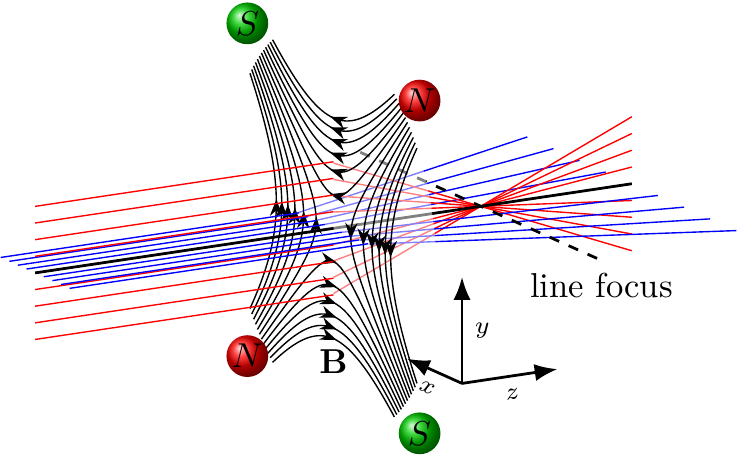}\caption{Paraxial optics of a thin quadrupole: Blue and red trajectories in
the two principal planes are focused and defocused, respectively.
Magnetic field lines illustrate the 45° rotation of the magnetic poles
with respect to the two principal planes. }
\end{figure}
Magnetic quadrupoles are used as stigmators and anisotropic focusing
elements in various CPO applications across the whole velocity range
\citep{Courant1952,Hawkes1970}. They focus in one plane and defocus
in the perpendicular one. The $\boldsymbol{B}$-field within the poles,
rotated about 45° with respect to the coordinate system, can be described
by linearly increasing Cartesian components (this expression is exact
for hyperbolic poles; otherwise it is a good approximation close to
the optical axis)

\begin{equation}
\boldsymbol{B}=-D_{2}\begin{pmatrix}y\\
x\\
0
\end{pmatrix}\,.\label{eq:B_quadrup}
\end{equation}
The vector potential in a Coulomb gauge ensuring z-independence reads
\begin{equation}
\boldsymbol{A}=\frac{1}{2}D_{2}\begin{pmatrix}0\\
0\\
x^{2}-y^{2}
\end{pmatrix}\,,
\end{equation}
which leads to a time independent Jacobian in the JVE (\ref{eq:Lie_coord})
\begin{equation}
\chi=\left(\begin{array}{c|c}
0 & \begin{array}{ccc}
\frac{1}{m} & 0 & 0\\
0 & \frac{1}{m} & 0\\
0 & 0 & \frac{1}{m}
\end{array}\\
\hline \begin{array}{ccc}
qD_{2}v & 0 & 0\\
0 & -qD_{2}v & 0\\
0 & 0 & 0
\end{array} & 0
\end{array}\right)\,.\label{eq:Lie_quadrup}
\end{equation}
and no further implications from the sharp cut-off fields. Accordingly,
the decoupling of the two principal planes is established at this
stage with the $z$-direction being a trivial propagation. The exactly
solvable transfer matrices for a quadrupole of length $l=vt$ in the
defocusing and focusing plane then read
\begin{align}
M_{x} & =\exp\begin{pmatrix}0 & \frac{l}{mv}\\
qD_{2}l & 0
\end{pmatrix}\\
 & =\begin{pmatrix}\mathrm{cosh}\left(kl\right) & \frac{1}{kp_{0}}\mathrm{sinh}\left(kl\right)\\
kp_{0}\mathrm{sinh}\left(kl\right) & \mathrm{cosh}\left(kl\right)
\end{pmatrix}\,,\nonumber 
\end{align}
\begin{equation}
M_{y}=\begin{pmatrix}\mathrm{cos}\left(kl\right) & \frac{1}{kp_{0}}\mathrm{sin}\left(kl\right)\\
-kp_{0}\mathrm{sin}\left(kl\right) & \mathrm{cos}\left(kl\right)
\end{pmatrix}\,,
\end{equation}
and
\begin{equation}
M_{z}=\begin{pmatrix}1 & \frac{l}{p_{0}}\\
0 & 1
\end{pmatrix}\,,
\end{equation}
with $k=\sqrt{\frac{qD_{2}}{p_{0}}}$. These transfer matrices need
not be corrected for the kinetic momentum since $A_{x}=A_{y}=0$ and
hence they are exactly equal to those obtained from the solution of
conventional paraxial theory \citep{Hawkes1966,Dragt1982}. 

The first order aberrations follow from 
\begin{equation}
\left(\frac{\partial X_{i}}{\partial x_{j}\partial x_{k}}\right)_{x\left(t\right)}=\frac{qD_{2}}{m}\left(\begin{array}{c|c}
\begin{array}{ccc}
-\delta_{i,3} & 0 & 0\\
0 & \delta_{i,3} & 0\\
0 & 0 & 0
\end{array} & \begin{array}{ccc}
0 & 0 & \delta_{i,4}\\
0 & 0 & -\delta_{i,5}\\
0 & 0 & 0
\end{array}\\
\hline \begin{array}{ccc}
0 & 0 & 0\\
0 & 0 & 0\\
\delta_{i,4} & -\delta_{i,5} & 0
\end{array} & 0
\end{array}\right)
\end{equation}
which leads to a second order transfer matrix $\mathcal{M}^{\left(2\right)}$
given by (\ref{eq:2nd_order_M}) with a few non-zero elements only.
In case of the transverse elements these pertain to dependencies on
the initial longitudinal momentum, i.e., 
\begin{align}
\mathcal{M}_{116}^{\left(2\right)}=\mathcal{M}_{161}^{\left(2\right)} & =\frac{kl\sinh\left(kl\right)}{4p_{0}}\\
\mathcal{M}_{146}^{\left(2\right)}=\mathcal{M}_{164}^{\left(2\right)} & =\frac{kl\cosh\left(kl\right)-\sinh\left(kl\right)}{4kp_{0}^{2}}\nonumber \\
\mathcal{M}_{226}^{\left(2\right)}=\mathcal{M}_{262}^{\left(2\right)} & =-\frac{kl\sin\left(kl\right)}{4p_{0}}\nonumber \\
\mathcal{M}_{256}^{\left(2\right)}=\mathcal{M}_{265}^{\left(2\right)} & =\frac{kl\cos\left(kl\right)-\sin\left(kl\right)}{4kp_{0}^{2}}\nonumber \\
\mathcal{M}_{416}^{\left(2\right)}=\mathcal{M}_{461}^{\left(2\right)} & =\frac{k\left(kl\cosh\left(kl\right)+\sinh\left(kl\right)\right)}{4}\nonumber \\
\mathcal{M}_{446}^{\left(2\right)}=\mathcal{M}_{264}^{\left(2\right)} & =\frac{kl\sinh\left(kl\right)}{4p_{0}}\nonumber \\
\mathcal{M}_{526}^{\left(2\right)}=\mathcal{M}_{562}^{\left(2\right)} & =-\frac{k\left(\sin\left(kl\right)+kl\cos\left(kl\right)\right)}{4}\nonumber \\
\mathcal{M}_{556}^{\left(2\right)}=\mathcal{M}_{565}^{\left(2\right)} & =-\frac{kl\sin\left(kl\right)}{4p_{0}}\nonumber 
\end{align}
The non-zero longitudinal deviation (i.e., $\mathcal{M}_{3..}^{\left(2\right)}$)
can be omitted because it introduces no effective second order deviation
in the image plane after time-lag correction. In order to obtain aberration
coefficients, i.e., the overall dependency of the final deviation
vector $\boldsymbol{Y}^{\left(2\right)}$ from (second order) polynomials
in the initial coordinates, we have to sum the symmetric contributions,
e.g., $C_{116}^{\left(2\right)}=M_{116}^{\left(2\right)}+M_{161}^{\left(2\right)}=2M_{116}^{\left(2\right)}$,
and add time lag aberrations (\ref{eq:2ndorder_timelag}) to final
plane perpendicular to optical axis 
\begin{equation}
Y_{i}{}^{\left(2t\right)}\left(t\right)=-\frac{1}{p_{0}^{2}}\left(\begin{array}{c|c}
0 & I_{3}\\
\hline 0 & 0
\end{array}\right)_{ij}\mathcal{M}_{jl_{1}}Y_{l_{1}}\left(0\right)Y_{6}\left(0\right)\,,
\end{equation}
which gives
\begin{align}
C_{116}^{\left(t\right)} & =-\frac{kl\sinh\left(kl\right)}{p_{0}}\\
C_{146}^{\left(t\right)} & =-\frac{l\cosh\left(kl\right)}{p_{0}^{2}}\nonumber \\
C_{226}^{\left(t\right)} & =\frac{kl}{p_{0}}\sin\left(kl\right)\nonumber \\
C_{256}^{\left(t\right)} & =-\frac{l}{p_{0}^{2}}\cos(kl)\nonumber \\
C_{366}^{\left(t\right)} & =-\frac{l}{p_{0}^{2}}\,.\nonumber 
\end{align}

Again, we omitted the non-zero longitudinal deviation. To obtain the
aberration coefficient, defined as the (second order) Taylor expansion
coefficient of the final coordinate from the starting conditions,
both contributions have to be summed up, e.g.,
\begin{equation}
C_{116}=C_{116}^{\left(2\right)}+C_{116}^{\left(t\right)}=-\frac{1}{2}\frac{kl\sinh\left(kl\right)}{p_{0}}=\frac{d^{2}Y_{1}\left(l\right)}{dp_{0}dY_{1}\left(0\right)}\,.
\end{equation}
Here the final result may be directly validated by direct verification
of the last equality. The other aberration coefficients read
\begin{align}
C_{146} & =\frac{-l\mathrm{cosh}\left(kl\right)-k^{-1}\mathrm{sinh}\left(kl\right)}{2p_{0}^{2}}\\
C_{226} & =\frac{kl\sin\left(kl\right)}{2p_{0}}\nonumber \\
C_{256} & =\frac{-l\cos\left(kl\right)-k^{-1}\sin\left(kl\right)}{2p_{0}^{2}}\nonumber \\
C_{416} & =\frac{k\left(kl\cosh\left(kl\right)+\sinh\left(kl\right)\right)}{2}\nonumber \\
C_{446} & =\frac{kl\sinh\left(kl\right)}{2p_{0}}\nonumber \\
C_{526} & =-\frac{k\left(\sin\left(kl\right)+kl\cos\left(kl\right)\right)}{2}\nonumber \\
C_{556} & =-\frac{kl\sin\left(kl\right)}{2p_{0}}\,,\nonumber 
\end{align}
 which correspond to the well-known chromatic aberrations of the magnetic
quadrupole (here we also noted deviations in the kinetic momentum,
i.e., $C_{4..}$ and $C_{5..},$ for completeness).

\subsection{Round Lens}

\begin{figure}
\includegraphics{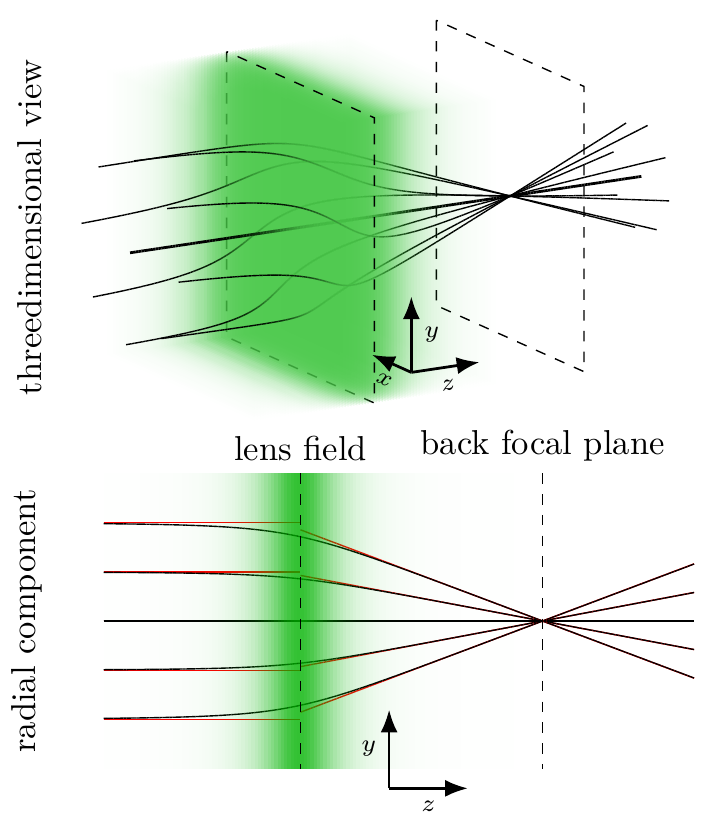}\caption{Paraxial optics of thin round lens: The trajectories are simultaneously
rotated and bend toward the optical axis. The former may be eliminated
by a transformation into a rotating frame of reference, the so called
Larmor reference frame. The black lines are analytical trajectories
through a Glaser bell-shaped field, the red lines those obtained from
the transfer matrix of a thin round lens as described in this section.}
\end{figure}

Round magnetic lenses are the principal focusing devices in electron
optical instruments with intermediate acceleration potential ($10$
kV - 1 MeV). They generate a confined magnetic field, which predominantly
points along the straight optical axis and has rotational symmetry.
Taking into account symmetry and neglecting radial dependency of the
axial field in the vicinity of the optical axis, the magnetic field
may be approximated with the help of Maxwell's law $\nabla\cdot\boldsymbol{B=0}$
in cylindrical coordinates

\begin{equation}
\frac{\partial}{\partial\rho}\left(\rho B_{\rho}\right)=-\rho\frac{\partial}{\partial z}B_{z}\rightarrow B_{\rho}=-\frac{\rho}{2}\frac{\partial}{\partial z}B_{z}
\end{equation}
or
\begin{equation}
\boldsymbol{B}=\underbrace{\begin{pmatrix}-\frac{\rho}{2}\frac{\partial}{\partial z}B_{z}\\
0\\
B_{z}
\end{pmatrix}}_{\textrm{cylindrical}}=\underbrace{\begin{pmatrix}-\frac{x}{2}\frac{\partial B_{z}}{\partial z}\\
-\frac{y}{2}\frac{\partial B_{z}}{\partial z}\\
B_{z}
\end{pmatrix}}_{\textrm{cartesian}}\,.
\end{equation}
We express the vector potentials in line gauge ensuring a vanishing
$z$-component 
\begin{equation}
\boldsymbol{A}=\begin{pmatrix}-\frac{zy}{2}\frac{\partial}{\partial z}\int\limits _{0}^{1}B_{z}\left(uz\right)udu-y\int\limits _{0}^{1}B_{z}(uz)udu\\
\frac{zx}{2}\frac{\partial}{\partial z}\int\limits _{0}^{1}B_{z}\left(uz\right)udu+x\int\limits _{0}^{1}B_{z}(uz)udu\\
0
\end{pmatrix}\,,
\end{equation}
which renders the JVE for the longitudinal component trivial. The
transverse components of the Jacobian of the Hamiltonian flow read
\begin{equation}
\chi_{xy}\left(t\right)=\frac{1}{m}\left(\begin{array}{cccc}
0 & \frac{1}{2}qB_{z} & 1 & 0\\
-\frac{1}{2}qB_{z} & 0 & 0 & 1\\
-\frac{1}{4}q^{2}B_{z}^{2} & 0 & 0 & \frac{1}{2}qB_{z}\\
0 & -\frac{1}{4}q^{2}B_{z}^{2} & -\frac{1}{2}qB_{z} & 0
\end{array}\right)\,.\label{eq:Lie_round_lens}
\end{equation}
The entries in the upper left and lower right $2\times2$ block indicate
that the round lens rotates the image by the Larmor frequency $\frac{qB_{z}}{2m}$.

In the next step we absorb the Larmor rotation into a rotating Larmor
frame of reference by employing the ansatz $\boldsymbol{Y}=O\widetilde{\boldsymbol{Y}}$with
the Larmor rotation matrix $O$. It follows that 
\begin{equation}
\frac{d\widetilde{\boldsymbol{Y}}}{dt}=\left(o^{T}+O^{T}\chi_{x\left(t\right)}O\right)\widetilde{\boldsymbol{Y}}
\end{equation}
with the infinitesimal rotation matrix
\begin{equation}
o=\frac{1}{m}\left(\begin{array}{cccc}
0 & \frac{1}{2}qB_{z} & 0 & 0\\
-\frac{1}{2}qB_{z} & 0 & 0 & 0\\
0 & 0 & 0 & \frac{1}{2}qB_{z}\\
0 & 0 & -\frac{1}{2}qB_{z} & 0
\end{array}\right)
\end{equation}
and $O=\exp\left(to\right)$. Consequently, the JVE in the Larmor
frame of reference has the Jacobian
\begin{align}
\widetilde{\chi}\left(t\right) & \coloneqq o^{T}+O^{T}\left(\frac{\partial X_{i}}{\partial x_{j}}\right)_{\boldsymbol{x}\left(t\right)}O\\
 & =\frac{1}{m}\left(\begin{array}{cccc}
0 & 0 & 1 & 0\\
0 & 0 & 0 & 1\\
-\frac{1}{4}q^{2}B_{z}^{2} & 0 & 0 & 0\\
0 & -\frac{1}{4}q^{2}B_{z}^{2} & 0 & 0
\end{array}\right)_{\boldsymbol{x}\left(t\right)}\,,\nonumber 
\end{align}
which is very similar to that of the quadrupole (\ref{eq:Lie_quadrup})
with the notable difference that a decoupled and identical focusing
action occurs in both transverse directions. As the matrix of the
system of first order differential equations depends on $t$ (via
$z$$\left(t\right)$) in this case, it is, however, not possible
to simply integrate the exponent before taking the matrix exponential
to obtain the exact result.

Nevertheless, increasingly good approximations, which preserve the
symplectic structure and hence phase space areas may be obtained with
the help of the Magnus expansion (\ref{eq:Magnus}). The first term
of which corresponds to the integration of the exponent, i.e., 
\begin{align}
\widetilde{M}_{xy}^{\left(I\right)} & =\exp\begin{pmatrix}0 & \frac{l}{mv}\\
-\frac{1}{4}\frac{q^{2}}{m}\int_{0}^{l/v}B_{z}^{2}dt & 0
\end{pmatrix}\label{eq:weak_lens}\\
 & =\left(\begin{array}{cc}
\cos\left(\varphi\right) & \frac{2\sqrt{l}}{q\sqrt{\int_{0}^{l}B_{z}^{2}dz}}\sin\left(\varphi\right)\\
-\frac{q\sqrt{\int_{0}^{l}B_{z}^{2}dz}}{2\sqrt{l}}\sin\left(\varphi\right) & \cos\left(\varphi\right)
\end{array}\right)\nonumber 
\end{align}
with 
\begin{equation}
\varphi\coloneqq\frac{q}{2mv}\sqrt{l\int_{0}^{l}B_{z}^{2}dz}\,.
\end{equation}
Note that we considered only one of the identical transverse directions
here for clarity. 

Similar to that of quadrupole, the chosen gauge implies that the kinetic
and canonical momentum coincide. The focal length for the weak lens
$\varphi\ll1$ then reads $\frac{1}{f}=\frac{1}{4}\frac{q^{2}}{\left(mv\right)^{2}}\int_{0}^{l}B_{z}^{2}dz$,
which corresponds exactly to famous Busch's formula for the focusing
action of a thin lens \citep{Busch(1926)}. Up to second order the
Magnus expansion (\ref{eq:Magnus}) involves a commutator 
\begin{equation}
\left[\widetilde{\chi}\left(t\right),\widetilde{\chi}\left(t_{1}\right)\right]=\frac{q^{2}\left(B_{z}^{2}\left(t\right)-B_{z}^{2}\left(t_{1}\right)\right)}{4m^{2}}\left(\begin{array}{cc}
1 & 0\\
0 & -1
\end{array}\right)
\end{equation}
of the Jacobian $\chi\left(t\right)$ matrix of the JVE (\ref{eq:Lie_coord})
and reads\begin{widetext}
\begin{align}
\widetilde{M}_{xy}^{\left(II\right)} & =\exp\left(\int_{0}^{l/v}dt\widetilde{\chi}_{ij}\left(t\right)+\frac{1}{2}\int_{0}^{l/v}dt\int_{0}^{t}dt_{1}\left[\widetilde{\chi}_{ij}\left(t\right),\widetilde{\chi}_{ij}\left(t_{1}\right)\right]\right)\\
 & =\exp\begin{pmatrix}\frac{q^{2}}{8m^{2}}\int_{0}^{l/v}\left(B_{z}^{2}t-\int_{0}^{t}B_{z}^{2}dt_{1}\right)dt & \frac{l}{mv}\\
-\frac{q^{2}}{4m}\int_{0}^{l/v}B_{z}^{2}dt & -\frac{q^{2}}{8m^{2}}\int_{0}^{l/v}\left(B_{z}^{2}t-\int_{0}^{t}B_{z}^{2}dt_{1}\right)dt
\end{pmatrix}\nonumber 
\end{align}
The latter may be further simplified by partial integration if the
initial and final plane considered are outside the field of the lens
\begin{align}
\widetilde{M}_{xy}^{\left(II\right)} & =\exp\begin{pmatrix}\frac{q^{2}}{4m^{2}}\int_{0}^{l/v}B_{z}^{2}tdt & \frac{l}{mv}\\
-\frac{q^{2}}{4m}\int_{0}^{l/v}B_{z}^{2}dt & -\frac{q^{2}}{4m^{2}}\int_{0}^{l/v}\int_{0}^{t}B_{z}^{2}dt
\end{pmatrix}\\
 & =\left(\begin{array}{cc}
\frac{q^{2}\int_{0}^{l}B_{z}^{2}zdz\,\sin\left(\varphi\right)}{4m^{2}v^{2}\varphi}+\cos\left(\varphi\right) & \frac{l\sin\left(\varphi\right)}{\gamma mv\varphi}\\
-\frac{q^{2}\int_{0}^{l}B_{z}^{2}dz\,\sin\left(\varphi\right)}{4mv\varphi} & \frac{-q^{2}\int_{0}^{l}B_{z}^{2}zdz\,\sin\left(\varphi\right)}{4m^{2}v^{2}\varphi}+\cos\left(\varphi\right)
\end{array}\right)\nonumber 
\end{align}
\end{widetext}with
\begin{equation}
\varphi\coloneqq\frac{q}{2m}\sqrt{\frac{l}{v^{2}}\int_{0}^{l}B_{z}^{2}dz-\frac{q^{2}}{4m^{2}v}\left(\int_{0}^{l}B_{z}^{2}dz\right)^{2}}\,.
\end{equation}
It is readily verified that ($\tilde{M}_{xy}^{\left(II\right)}$)
is again symplectic. A detailed comparison of such higher-order Magnus
expansions with paraxial trajectories obtained by numerical step solvers
(or exact solutions of the Glaser field) is warranted to beyond the
scope of this paper and will be conducted elsewhere. Depending on
the required accuracy, such explicit transfer matrix expressions may
be used to derive more accurate expressions of aberrations integrals
of round lenses. Similarly improved closed expression can be also
derived for the inhomogeneous quadrupole (e.g., including fringing
fields) or the following devices.

\subsection{Wien Filter}

\begin{figure}
\includegraphics[width=1\columnwidth]{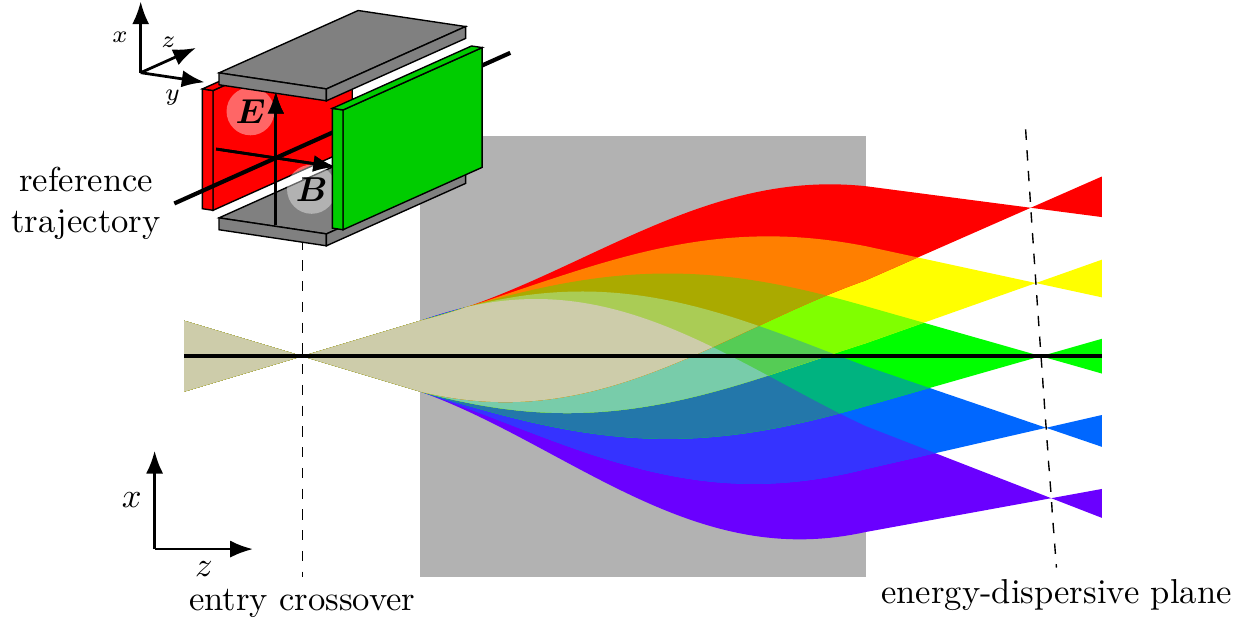}\caption{\label{fig:Wien_filter}Paraxial optics of a Wien filter. As the longitudinal
deviations are not causing any additional transverse deflections,
they are not shown.}
\end{figure}
CPO devices with non-zero transverse magnetic and electrostatic fields
along the optical axis are used predominantly as deflectors and spectrometers.
The Wien filter is such a device that employs crossed electric and
magnetic fields to separate charged particles according to their velocity
(in case of charged particles of the same $q/m$ ratio) or mass (in
case of ions of defined energy with different $q/m$ ratio). It has
the unique property (among velocity filters) of a straight reference
trajectory, i.e., the optical axis $x$ (see Fig. \ref{fig:Wien_filter}).
In the following, we will again consider only the homogeneous case
(i.e., the inner region of the filter), which contains only constant
electric and magnetic fields
\begin{equation}
\boldsymbol{B}=\begin{pmatrix}0\\
B_{y}\\
0
\end{pmatrix}\,,\boldsymbol{E}=\begin{pmatrix}E_{x}=vB_{y}\\
0\\
0
\end{pmatrix}
\end{equation}
which are perpendicular and obey a fixed relationship ensuring a straight
reference trajectory. The vector potential in a convenient gauge $(A_{x}=A_{y}=0)$
reads
\begin{equation}
\mathbf{A}(\mathbf{r})=\begin{pmatrix}0\\
0\\
-B_{y}x
\end{pmatrix}\,.
\end{equation}
yielding the following Hamilton flow Jacobian (\ref{eq:Lie_coord})
for the deviations in $x$- and $z$-direction ($y$ is a trivial
drift)
\begin{equation}
\chi_{xz}=\frac{1}{m}\left(\begin{array}{cccc}
0 & -qB_{y} & 1 & 0\\
0 & 0 & 0 & 1\\
0 & 0 & 0 & 0\\
0 & -q^{2}B_{y}^{2} & qB_{y} & 0
\end{array}\right)\,.
\end{equation}
In contrast to the quadrupole, the sharp cut-off field additionally
modifies the paraxial trajectory in phase space by introducing a sharp
jump in the transverse momentum at the entrance and exit plane. Indeed,
the latter corresponds to the gauge transformation in the initial
and final plane discussed further below. As the Hamilton flow Jacobian
is time-invariant, the transfer matrices for a Wien filter of length
$l$ can be obtained easily by integrating the Jacobian $\boldsymbol{\chi}$
and taking the matrix exponential
\begin{align}
\mathcal{M}_{xz} & =\exp\left(\begin{array}{cccc}
0 & 0 & \frac{l}{mv} & 0\\
qB_{y}\frac{l}{mv} & 0 & 0 & \frac{l}{mv}\\
-q^{2}B_{y}^{2}\frac{l}{mv} & 0 & 0 & -qB_{y}\frac{l}{mv}\\
0 & 0 & 0 & 0
\end{array}\right)\\
 & =\left(\begin{array}{cccc}
\cos\varphi & 0 & \frac{\sin\varphi}{qB_{y}} & \frac{\cos\varphi-1}{qB_{y}}\\
\sin\varphi & 1 & -\frac{\cos\varphi-1}{qB_{y}} & \frac{\sin\varphi}{qB_{y}}\\
-qB_{y}\sin\varphi & 0 & \cos\varphi & -\sin\varphi\\
0 & 0 & 0 & 1
\end{array}\right)\nonumber 
\end{align}
with $\varphi\coloneqq qB_{y}\frac{l}{mv}$. To obtain the gauge-independent
kinetic momentum transfer we can exploit that the gauge linearly depends
on the spatial coordinates, which relates the canonical and kinetic
transfer matrices by a similarity transformation (\ref{eq:sim_trafo}),
yielding
\begin{align}
M_{xz} & =\left(\begin{array}{cccc}
1 & 0 & 0 & 0\\
0 & 1 & 0 & 0\\
0 & 0 & 1 & 0\\
qB_{y} & 0 & 0 & 1
\end{array}\right)\mathcal{M}_{xz}\left(\begin{array}{cccc}
1 & 0 & 0 & 0\\
0 & 1 & 0 & 0\\
0 & 0 & 1 & 0\\
-qB_{y} & 0 & 0 & 1
\end{array}\right)\\
 & =\left(\begin{array}{cccc}
1 & 0 & \frac{\sin\varphi}{qB_{y}} & \frac{\cos\varphi-1}{qB_{y}}\\
0 & 1 & -\frac{\cos\varphi-1}{qB_{y}} & \frac{\sin\varphi}{qB_{y}}\\
0 & 0 & \cos\varphi & -\sin\varphi\\
0 & 0 & \sin\varphi & \cos\varphi
\end{array}\right)\,,\nonumber 
\end{align}
and 
\begin{equation}
M_{y}=\begin{pmatrix}1 & \frac{l}{p_{0}}\\
0 & 1
\end{pmatrix}\,,
\end{equation}
This result coincides with that obtained from the Newtonian equations
of motion. Note that the lower left sector of $M_{xz}$ is empty,
which means that the Wien filter acts similar to a telescope. Notwithstanding,
a point on the optical axis is imaged into a point, with the position
of the focal point depending on the initial longitudinal momentum
(see Fig. \ref{fig:Wien_filter}), which is the determining property
in energy filter and spectrometer applications. We finally note that
initial momentum deviations also entail longitudinal deviations in
the final plane. These may be, however, neglected in the paraxial
limit because the pertaining additional propagation along the straight
optical axis do not induce any additional transversal deviations (see
below for another example, where this is not the case).

The first order aberrations are determined completely by time lag
aberrations (\ref{eq:2ndorder_timelag}) because
\begin{equation}
\left(\frac{\partial X_{i}}{\partial x_{j}\partial x_{k}}\right)_{x\left(t\right)}=0\,.
\end{equation}
Taking into account gauge, we have to evaluate
\begin{equation}
M_{il_{1}l_{2}}^{\left(2t\right)}=G_{ii_{1}}\left(t\right)\mathcal{M}_{i_{1}j_{1}j_{2}}^{\left(2t\right)}G_{j_{1}l_{1}}^{-1}\left(0\right)G_{j_{1}l_{1}}^{-1}\left(0\right)
\end{equation}
yielding
\begin{align}
C_{144}	&=-\frac{\cos\varphi\left(\cos\varphi-1\right)}{qB_{y}p_{0}} \nonumber \\
C_{146}	&=-\frac{\sin\varphi\left(2\cos\varphi-1\right)}{qB_{y}p_{0}} \nonumber \\
C_{166}	&=\frac{\sin^{2}\varphi}{qB_{y}p_{0}} \nonumber \\
C_{254}	&=\frac{\cos\varphi-1}{qB_{y}p_{0}} \nonumber \\
C_{256}	&=-\frac{\sin\varphi}{qB_{y}p_{0}} \nonumber \\
\end{align}
These are the aberrations of the Wien filter with sharp cut-off fields.
For a comprehensive analytic treatment including higher orders and
fringing fields see Ref. \citep{Tsuno2013}.

\subsection{Sector Magnet\label{subsec:Sector-Magnet}}

\begin{figure}
\includegraphics[width=1\columnwidth]{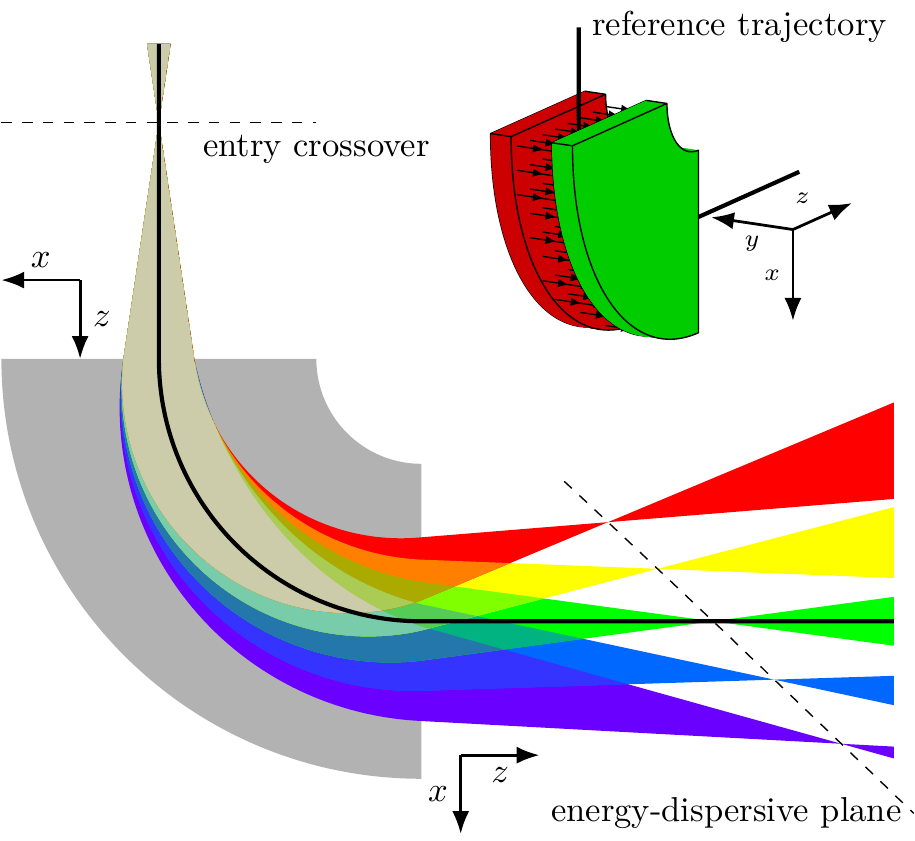}\caption{\label{fig:Sector-magnet}Paraxial optics of a sector magnet. Note
that the coordinate system is turned alongside the reference trajectory
after the exit plane.}
\end{figure}
Sector magnets refer to opposing magnetic pole piece geometries being
limited by entrance and exit planes such to create a sector of a circle
(see Fig. \ref{fig:Sector-magnet}). They find application in spectrometers,
energy filters, particle separators as well as beam guiding systems.
Considering again the case of homogeneous fields only (i.e., parallel
pole pieces with sharp cut-off), we can directly copy the transfer
matrix of the Wien filter in the $x-z$ plane

\begin{equation}
M=\left(\begin{array}{cccc}
1 & 0 & \frac{\sin\varphi}{qB_{z}} & \frac{\cos\varphi-1}{qB_{z}}\\
0 & 1 & -\frac{\cos\varphi-1}{qB_{z}} & \frac{\sin\varphi}{qB_{z}}\\
0 & 0 & \cos\varphi & -\sin\varphi\\
0 & 0 & \sin\varphi & \cos\varphi
\end{array}\right)\,.
\end{equation}

We again note the missing dependency of the final kinetic momentum
on the initial position, which, contrary to the Wien filter, however,
does not imply a missing focusing action, i.e., an induced convergence
of parallel trajectories, in the sector magnet. The latter is induced
by considering that longitudinal deviations (see Fig. \ref{fig:Sector-magnet})
effectively introduce an additional deflection due to additional propagation
lengthsdepending linearly on the initial beam position. To illustrate
that point we first compute the time of flight $\delta t$ required
for the particle to cover that longitudinal shift by projecting the
spacial part of the deviation vector on the tangent of the optical
axis
\begin{equation}
-\sin\varphi X_{1}\left(t\right)+\cos\varphi X_{3}\left(t\right)\approx\omega R\delta t\,,
\end{equation}
and hence 
\begin{align}
\delta t & \approx\frac{-\sin\varphi X_{1}\left(t\right)+\cos\varphi X_{3}\left(t\right)}{\omega R}\,.
\end{align}
Adding the ensuing additional propagation along the optical axis to
the particle we obtain the first order additional deviation of the
beam 
\begin{align}
\boldsymbol{X}'\left(t\right) & =\omega\left(\begin{array}{cccc}
0 & -\delta t & 0 & 0\\
\delta t & 0 & 0 & 0\\
0 & 0 & 0 & -\delta t\\
0 & 0 & \delta t & 0
\end{array}\right)\boldsymbol{x}\left(t\right)\\
 & =R\omega\left(\begin{array}{cccc}
0 & -\delta t & 0 & 0\\
\delta t & 0 & 0 & 0\\
0 & 0 & 0 & -\delta t\\
0 & 0 & \delta t & 0
\end{array}\right)\begin{pmatrix}\cos\varphi\\
\sin\varphi\\
-\gamma m\omega\sin\varphi\\
\gamma m\omega\cos\varphi
\end{pmatrix}\nonumber \\
 & =R\omega\begin{pmatrix}-\sin\varphi\\
\cos\varphi\\
-qB_{z}\cos\varphi\\
-qB_{z}\sin\varphi
\end{pmatrix}\delta t\nonumber \\
 & =\left(\begin{array}{cccc}
\sin^{2}\varphi & -\sin\varphi\cos\varphi & 0 & 0\\
-\cos\varphi\sin\varphi & \cos^{2}\varphi & 0 & 0\\
qB_{z}\cos\varphi\sin\varphi & -qB_{z}\cos^{2}\varphi & 0 & 0\\
qB_{z}\sin^{2}\varphi & -qB_{z}\cos\varphi\sin\varphi & 0 & 0
\end{array}\right)\boldsymbol{X}\left(t\right)\nonumber \\
 & =M'\boldsymbol{X}\left(0\right)\nonumber 
\end{align}
with 
\begin{equation}
M'=\left(\begin{array}{cccc}
\sin^{2}\varphi & -\sin\varphi\cos\varphi & 0 & 0\\
-\cos\varphi\sin\varphi & \cos^{2}\varphi & 0 & 0\\
qB_{z}\cos\varphi\sin\varphi & -qB_{z}\cos^{2}\varphi & 0 & 0\\
qB_{z}\sin^{2}\varphi & -qB_{z}\cos\varphi\sin\varphi & 0 & 0
\end{array}\right)M\,.
\end{equation}
The additional deviation $\boldsymbol{X}'\left(t\right)$ needs to
be added to the final $\boldsymbol{X}\left(t\right)$ to obtain the
correct deviation in the image plane, which can be alternatively written
as $\boldsymbol{X}'\left(t\right)+\boldsymbol{X}\left(t\right)=\mathscr{M}\boldsymbol{X}\left(0\right)$
defining a modified transfer matrix 
\begin{align}
\mathscr{\mathscr{M}} & \coloneqq O\left(M+M'\right)\\
 & =\left(\begin{array}{cccc}
\cos\varphi & \sin\varphi & 0 & 0\\
-\sin\varphi & \cos\varphi & 0 & 0\\
0 & 0 & \cos\varphi & \sin\varphi\\
0 & 0 & -\sin\varphi & \cos\varphi
\end{array}\right)\left(M+M'\right)\nonumber \\
 & =\left(\begin{array}{cccc}
\cos\varphi & \sin\varphi & \frac{\sin\varphi}{qB_{z}} & \frac{1-\cos\varphi}{qB_{z}}\\
0 & 0 & 0 & 0\\
-qB_{z}\sin\varphi & qB_{z}\cos\varphi & \cos\varphi & \sin\varphi\\
0 & 0 & 0 & 1
\end{array}\right)\nonumber 
\end{align}
Here we have incorporated an additional rotation $O$ transforming
the vector in laboratory frame into a frame rotated with the optical
axis. The obtained $\mathscr{\mathscr{M}}$ is equivalent to the classical
result for the sector magnet \citep{Enge1967,Wollnik1987,Enge1967a}
and exhibits the familiar focusing effect in radial direction (see
lower left sector of $\mathscr{\mathscr{M}}$) including Barber's
rule \citep{Barber1933}.
\end{document}